\documentclass{aa}  
\input{styleinuse}
\usepackage{subfig}
\usepackage{soul}	
\usepackage{xcolor}

\begin{document}

   \title{The galactic bubbles of starburst galaxies}

   \subtitle{The influence of  galactic large-scale magnetic fields}
\titlerunning{Magnetize galactic bubble}
\authorrunning{Meliani et al.} 
   \author{Z. Meliani\inst{1}, P. Cristofari\inst{1}, A. Rodr\'iguez-Gonz\'alez\inst{2}, G. Fichet de Clairfontaine\inst{3},  E. Proust\inst{4}, E. Peretti\inst{5}\inst{6}
          }

   \institute{$^{1}$Laboratoire Univers et Théories, Observatoire de Paris, Université PSL, Université Paris Cité, CNrs, F-92190 Meudon, France
 \\
   $^{2}$Instituto de Ciencias Nucleares, Universidad Nacional Aut\'onoma de M\'exico, Ap. 70-543, 04510 CDMX, M\'exico \\
    $^{3}$Julius-Maximilians-Universität Würzburg, Fakultät für Physik und
    Astronomie, Emil-Fischer-Str. 31, D-97074 Würzburg, Germany \\
     $^{4}$Ecole centralesupelec, Universit\'e Paris-Saclay, 91190 Gif-sur-Yvette, France\\
     $^{5}$Niels Bohr International Academy, Niels Bohr Institute,University of Copenhagen, Blegdamsvej 17, DK-2100 Copenhagen, Denmark \\
     $^{6}$Université Paris Cité, CNRS, Astroparticule et Cosmologie, 10 Rue Alice Domon et Léonie Duquet, F-75013 Paris, France
     }

   \date{Received September 15, 1996; accepted March 16, 1997}

 
  \abstract
{The galactic winds of starburst galaxies (SBGs) give rise to remarkable structures on kiloparsec scales. However, the evolution and  shape of these giant wind bubbles, as well as the properties of the shocks they develop, are not yet fully understood.}
{We aim to understand what shapes the galactic winds of SBGs,  with a particular focus on  the role of large-scale magnetic fields in the dynamical evolution of galactic wind-inflated bubbles. In addition, we aim to explore where the conditions for efficient particle acceleration are met in these systems.}
{We performed magnetohydrodynamic simulations with the AMRVAC code (Adaptive Mesh Refinement Versatile Advection Code) with various configurations of the galactic medium density profile and magnetization.}
{We observe that the large-scale magnetic field, in which galactic winds expand, can impact the structure and evolution of inflated bubbles. However, the typical structures observed in starburst galaxies, such as M82, cannot be solely explained by the  magnetic field structures that have been considered. This highlights the importance of other factors, such as the galactic disk, in shaping the galactic bubble. Furthermore, in all the magnetized cases we investigated, the forward wave resulting from the expanding bubbles only results in compression waves, whereas the wind termination shock features high Mach numbers, making it a promising site for diffusive shock acceleration up to $\sim 10^2 \rm PeV$. The synthetic X-ray images generated from our models reveal an envelope surrounding the bubbles that extends up to 2 kpc, which could correspond to the polarized emission observed from planar geometry in M82, as well as a large structure inside the bubble corresponding to the shocked galactic wind. Additionally, our findings indicate that, as observed with the SOFIA instrument, a large ordered magnetic field is associated with the free galactic wind, while a more turbulent magnetic field is present in the shocked region.}
   {}
   
   \keywords{Galaxies: starburst --
              Galaxies: magnetic fields --Galaxy: disk-- Galaxy: halo
                 -- Stars: mass-loss
                ,Magnetohydrodynamics --
              Methods: numerical }

   \maketitle
  
%

\section{Introduction}

Starburst galaxies (SBGs) can be characterized as galaxies with an intense star formation rate, which consequently leads to a high rate of supernova (SN) explosions. This intense star formation activity often takes place in a relatively  compact volume (typically with a size of a few to a hundred parsecs), referred to as a starburst nucleus~\citep[SBN,][]{kennicutt1998}. 
The extreme amount of energy released by star formation and SNe in the compact SBN environment can power large-scale outflows~\citep[][]{Chevalier1985}.
Indeed, on kiloparsec scales, large structures influenced by the activity of the SBN are often observed \citep[see][and references therein]{Veilleux_etal_2005ARA&A..43..769V}.

Starburst galaxies are of considerable interest for several reasons, particularly due to their demonstrated ability to accelerate particles to very-high-energy (VHE) domains. The detection of VHE $\gamma$-ray emissions from M82 by Fermi LAT at GeV energies~\citep{M82_Fermi} and by Veritas at energies exceeding 700 GeV~\citep{M82_Veritas}, as well as from NGC 253 at energies exceeding 220 GeV~\citep{Abdalla_Etal_2018A&A...617A..73H}, provides direct evidence of efficient particle acceleration in these systems. Particle acceleration is anticipated to take place in SBGs \citep{Yoast-Hull2013,peretti2019,Krumholz2020}, and potentially within the shocks generated by the large-scale bubbles produced by SBG winds \citep{anchordoqui1999,anchordoqui2018,Romero2018,Muller2020,Peretti2022}.
In addition, recent anisotropy measurements of the arrival direction of the highest-energy cosmic rays \citep{Auger_anisotropy} have highlighted SBGs as one of the most promising source classes of EeV particles, thereby stimulating detailed studies on the effect of the starburst environment on ultra-high-energy particles \citep[see e.g.,][and references therein]{Condorelli2023}.

Starburst winds are believed to be driven by the mechanical and thermal pressure from SNe and stars~\citep{zhang2018}. 
The evolution of these structures, on typical timescales of the order of  $\sim 1-10^2$ Myr, is complex, but  in general the galactic wind leads to the inflation of  thermally driven large-scale bubbles. 
The evolution of these wind bubbles is somewhat analogous to stellar bubbles, but on galactic scales. 
The wind bubble dynamics have been studied analytically~\citep{castor1975,weaver1977,koo1992} and through simulations~\citep{Tenorio-Tagle_1997,Strickland-Stevens-2000,Strickland_2002,Melioli_2013,Fielding_2017,Schneider_2020,Owen_Ellis,Nguyen2022}, but several aspects of the dynamical evolution still remain unclear~\citep{lancaster2021,lancaster2021_2}, especially the influence of the galactic disk and the large-scale magnetic field.

The formation of galactic winds is closely tied to the winds originating from massive stellar clusters situated near the galactic center. These stellar clusters undergo evolution within the galactic disk. 
Research conducted by  \citet{1974MNRAS.169..395B}; \citet{ary2008}; and \citet{ary2009} emphasizes the significance of galactic disk stratification in terms of both accelerating and propagating the galactic wind within it.

As previously mentioned, the magnetic field represents another essential physical component observed within SBGs. Its geometry and strength in SBGs have been investigated using various observing techniques. Nonthermal emission in radio, observed with the Very Large Array (VLA) \citep{Reuter_1994A&A...282..724R}, provides evidence of a large-scale poloidal magnetic field structure in M82. Analyses of emission and polarization at 18 and 22 cm reveal a structured planar geometry in the galactic center that extends up to 2 kpc \citep{Adebahr_etal_2017A&A...608A..29A}. Recent observations \citep{lopezrodriguez2021} of thermal polarized emission with the SOFIA instrument show that the magnetic field is a combination of a large-scale ordered potential field that should be associated with the galactic wind, as well as a small-scale turbulent field associated with bow-shock-like features.
The important role of magnetic fields in the dynamical evolution of galaxies has been discussed in several studies~\citep{chyzy2006,beck2015,chyzy2016,krause2019}. However as mentioned earlier, a clear understanding of the formation of the large-scale structures is still missing.

The magnetic field also plays a crucial role in influencing the strength, the shape, and the stability of shocks~\citep[e.g.,][]{vanmarle2015, Meyer_2022MNRAS.515L..29M,Meyer_etal_2023MNRAS.tmp.3366M}. In fact, the magnetic field can deflect the galactic wind, weaken shocks, and suppress certain types of instabilities such as Rayleigh-Taylor instabilities~\citep[e.g.,][]{vanMarle_etal_2014A&A...561A.152V} and Kelvin-Helmholtz instabilities~\citep[e.g.,][]{Keppens_etal_1999JPlPh..61....1K}. Additionally, it can induce new types of instabilities. The influence of the magnetic field on the flow and instabilities depends not only on the strength of the magnetic field but also on the angle between the magnetic field and the flow direction.

In the case of winds from massive stars, the importance of magnetic fields on the shaping of the wind bubbles has been shown~\citep{vanmarle2015}. Similarly, on larger galactic scales, the interstellar medium (ISM) magnetic fields can also be of importance in the shaping of the wind bubble.
In some cases, such as  in M82, observations from the radio to the far--infrared domain have helped to understand the magnetic field structure on larger scales~\citep{reuter1992,Reuter_1994A&A...282..724R,adebahr2013,Adebahr_etal_2017A&A...608A..29A,harper2018,lopezrodriguez2021}. 
Understanding the link between the magnetic field structure and the galactic wind bubble is important in the general context of galaxy formation, and of the study of starburst galaxies, but also in the context of the origin of cosmic rays. Indeed, the geometry of the problem can potentially affect particle acceleration from SBG winds, and particle escape from SBNi.  

%
%

Despite the significant scientific progress made in recent decades, several important aspects still require further investigation. One of these aspects is the deformation and deviation of wind bubbles from the neat spherical geometry often assumed to describe them. The effect of the halo medium on their evolution is another area of interest. The efficiency of shocks in these bubbles as particle accelerators is also a topic of ongoing research. The maximum energy achievable in starburst-driven winds remains to be determined. The role of magnetic fields in shaping these bubbles is another important area of study. Finally, understanding how these bubbles in turn affect the magnetic field structures is also of paramount importance.


By means of magnetohydrodynamic (MHD) simulations using the AMRVAC code by \citet{Keppens_etal_2021}, we study the evolution of large-scale bubbles inflated by galactic winds of SBGs. We particularly focus on the case of M82, one of the most well-monitored SBGs, and adopt physical parameters compatible with what has been inferred from observations.

In this work, we begin by introducing the density and pressure of the galactic medium, as well as the characteristics of the galactic wind that we use in our study of M82 in Section 2. In Section 3, we describe the numerical method that we employed. In Section 4, we discuss the physics of the wind bubble and its evolution. Finally, in Section 5, we analyze the results of the simulations with respect to various profiles of galactic medium density, pressure, and magnetization.

\section{The physics of the wind bubble}\label{Sec:The physics of the wind bubble}


The injection of gas and energy from galactic winds and/or supernova explosions, whether from individual stars or from young stellar clusters, results in the formation of hot gas bubbles that expand due to the pressure difference with their surrounding environment. The evolution of these small-scale bubbles can be approximated as an expanding sphere \citep{Chevalier1985}, but this approximation becomes inadequate when their size exceeds the scale of variation of the medium in which they propagate. 
The physics of the galactic wind launched from the central nucleus is similar to the wind produced by massive stars or stellar clusters \citep{weaver1977} in terms of its basic properties and behavior. 
In particular, the outflow initially expands with constant velocity (free expansion phase). 
The galactic wind, being supersonic, develops a forward shock, that precedes it in the external medium while, at the same time, the collision of the wind material with the external matter results in the formation of a wind termination shock. 
The latter is geometrically oriented toward the central engine, and it is sometimes referred to as reverse shock. 
The external material reached by the forward shock is swept up, heated and accumulated in a layer physically separated from the wind material by the contact discontinuity. 
During the free expansion phase, the two shocks proceed at constant speed, very close to each other, until the swept up mass becomes comparable with the total mass of the wind. 
As soon as this happens, the system enters the deceleration phase during which analytic models \citep{koo1992} predict the forward shock to expand as $R_{\rm fs}\sim t^{3/5}$ while the wind termination shock evolves as $R_{\rm ts}\sim t^{2/5}$. 
At a later time, if the pressure in the shocked wind becomes comparable with the pressure of the external medium, the wind bubble might enter the pressure-confined stage during which the wind termination shock could stall. While the pressure-confined stage can typically happen at a later time for the portion of the outflow expanding in the galactic halo, it could be reached quickly for the fraction of the wind expanding in the galactic disk.

The wind outflowing from the central nucleus has a velocity of $v_{\rm w}$ and a mass-loss rate of $\dot{M}_{\rm w}$, which leads to the following density profile just outside the SBN:

\begin{equation}\label{Eq:wind_density}
    \rho_{\rm w} (r)= \frac{\dot{M}_{\rm w}}{4 \pi \,v_{\rm w} \,r^2}\,.
\end{equation}
The dense supersonic wind generates shock wave that propagates through the interstellar medium, leading to a significant temperature increase given by the equation,


\begin{equation}\label{Eq:T_fwd_shock}
    T_{\rm{fs}}=1.4\times 10^7 \,\left (\frac{v_{\rm{fs}}}{1000{ \rm \,km\,s^{-1}} } \right)^2 \,\mathrm{K}\,,
\end{equation}
where $v_{\rm fs}$ is the forward shock wave velocity in km\,s$^{-1}$.


Different from the standard wind bubble theory, the ISM density gradient in disk galaxies can cause galactic bubbles to expand faster in the direction of the density gradient, similar to the formation of Laval nozzles. 
When the density gradient is present, the propagation velocity of the bubble in that direction can exceed the escape velocity,
particularly when the ambient density and pressure are low at high altitudes.
This can result in large Mach numbers at the forward shock in this region, which sweep up galactic material, forming a shell-like structure. 
On the other hand, in the direction perpendicular to the galactic axis, the lateral expansion of the bubble occurs at velocities that are mainly sonic, resulting in regions of compressed gas with much lower Mach numbers. 
The wind termination shock is reached by a very fast wind which has cooled adiabatically. Therefore, the resulting Mach number can be as high as several tens.
The interior of the shocked wind bubble contains gas that has been injected by the galactic winds and supernova remnants from the central stars cluster at galactic center. After being shocked by the termination shock, it can reach a temperature of the order of millions of degrees.
Radiative cooling on such hot gas is less efficient due to the relatively low density of the halo.

Large-scale magnetic fields can play a significant role in the star formation process and possibly also in the expansion of starburst-driven galactic bubbles. Observations of M82 have revealed the presence of large-scale and ordered magnetic fields. The presence of these magnetic fields can modify the properties of shocks, which in turn can affect the characteristics of the resulting bubbles.
If magnetic fields are present, the nature of shocks is no longer determined only by the speed and sound speed of the wind. Instead, shock strength also depends on the magnetosonic speed, particularly the fast magnetosonic speed. In fact, if the speed of the shock drops below the fast magnetosonic speed, the nature of the shock can change from J-type (strong-jump) to C-type (continuous variation as a compression wave). This change in shock nature can occur when the fast magnetosonic speed becomes significant and carries information ahead of the shock. Furthermore, the decrease in shock intensity and the resulting reduction in density, pressure, and velocity jump can have a substantial impact on particle acceleration, impeding the effective injection of thermal particles into the diffusive shock acceleration mechanism.
The impact of magnetic fields on the conditions of shock jumps also depends on their relative orientation. The angle between the direction normal to the shock wave surface and the direction of the magnetic field can alter the characteristic speeds, such as the fast and slow magnetosonic speeds, and the Alfvén speed.


During the initial phase of a galactic bubble’s expansion, a uniform magnetic field primarily affects the forward shock, which propagates through the external medium. This continues until the magnetic pressure within the compressed, swept-up shell becomes comparable to the wind ram pressure, as described by Equation 7 in \cite{vanmarle2015}. Consequently, the expansion of the shell perpendicular to the magnetic field stops when it reaches a critical radius $R_{\rm B}$, which is given by:

\begin{equation}\label{Eq:Critical_radius_B}
R_{\rm B} = \frac{\sqrt{2 \, \dot{M}_{\rm w} \, v_{\rm w}}}{B} \,,
\end{equation}

where $\dot{M}_{\rm w}$ represents the mass flux of the wind from the central source, $v_{\rm w}$ denotes the terminal wind speed, and $B$ signifies the strength of the magnetic field parallel to the shock surface.

On the other hand, in the case of a bubble expanding in a dipolar magnetic field, the influence of the magnetic field decreases with distance, causing its strength to vary as $\sin{\theta}/R^3$. Consequently, the intensity of the shock increases as the bubble expands and more over critical radius $R_{\rm B}$ depends in the polar angle $\theta$.  
\section{Model}
In this paper, we aim to study the influence of galactic medium properties, such as density, pressure, and magnetization, on the evolution of galactic bubbles in M82 (NGC 3034), widely recognized as the archetypal starburst galaxy \citep[e.g.,][]{Rieke_etal_1980ApJ...238...24R,Strickland-Stevens-2000}. Located at a distance of  $\approx 3.6 \mathrm{Mpc}$ \citep{Freedman_1994ApJ...427..628F,Dalcanton_2009ApJS..183...67D,Vacca_etal_2015ApJ...804...66V}, it is nearly edge-on, with an inclination of approximately $80^\circ$ \citep{McKeith_etal_1995A&A...293..703M}, making it an ideal candidate for the study of biconical galactic outflows along its axis. The current starburst activity, ongoing for ($\approx\,10\,\mathrm{Myr}$), is centered on the galactic nucleus and spans a diameter of around $500 \,\mathrm{pc}$.

Our investigation encompasses various scenarios for the physical environment in which the starburst wind originates and evolves. These scenarios include different halo density and pressure distributions, as well as combinations of halo and disk density and pressure distributions. We also examine different configurations of the magnetic field, specifically considering three types: a uniform magnetic field aligned with the galaxy's axis, a dipolar magnetic field, and cases involving a superposition of uniform and dipole magnetic fields. Detailed descriptions of these cases can be found in Table~\ref{tab:init_cond}.

\subsection{Galactic wind}
Energy-driven galactic winds, powered by massive-star supernovae and stellar winds, are imposed within the inner radial boundary at $R_{\rm wind, \mathrm{boundary}} = 130 \, \mathrm{pc}$ as an isotropic supersonic wind with a velocity of $v_{\rm wind} = 1000 \, \text{km/s}$ and a Mach number of approximately $\mathcal{M} = 1.01$ which correspond to a temperature of approximately $T_{\rm wind}\sim 5\times 10^7 \,\mathrm{K}$. As a result, the wind is thermally accelerated to reach its terminal speed of approximately $v_{\rm wind} \approx 1850 \, \text{km/s}$ and a Mach number $\mathcal{M}_{\rm wind}\,\sim\,10$ at a distance of around $500 \, \text{pc}$. This terminal wind speed is consistent with the work of \citet{Melioli_2013, Chevalier1985, Veilleux_etal_2005ARA&A..43..769V,Strickland_Heckman_2007ApJ...658..258S,Bustard16,Nguyen_Thompson_10.1093/mnras/stab2910}. It is worth noting that setting a Mach number $\mathcal{M} \ge 20$ with a wind velocity of $v_{\rm wind} \approx 2000$, or setting $\mathcal{M} \approx 1.01$ with a wind velocity of $v_{\rm wind} \approx 1000$, results in the same type of dynamics and shape for the bubble.

Given the focus of this paper on the interaction between the wind and the galactic medium, which primarily depends on the total momentum of the wind, we position the supersonic wind closer to the galactic center, resulting quasi-equipartition between the thermal and kinetic momentum at inner boundary. However, it is noteworthy that the galactic wind could reach its final speed of $v_{\rm wind} > 1800 \,\mathrm{km/s}$ at a further distance of $2\,\mathrm{kpc}$, as it may continue to accelerate due to the influence of thermal pressure, radiation pressure, and cosmic ray pressure, as discussed in \citep[e.g.,][]{Yu_etal_2020MNRAS.492.3179Y,Nguyen_Thompson_10.1093/mnras/stab2910}. Additionally, the dissipation of Alfvén waves could contribute to its acceleration \citep{Shimoda_Inutsuk_2022ApJ...926....8S}. However, for the scale of several kiloparsecs relevant to our  study, as demonstrated by \citet{Yu_etal_2020MNRAS.492.3179Y}, thermal pressure is more effective in accelerating a wind, followed by cosmic ray pressure and then radiation pressure on dust  \citep[e.g.,][]{Thompson_etal_2015MNRAS.449..147T}. In all the cases investigated here, only the effects of thermal energy and kinetic energy   are taken into account.

For the simulations perfomed in this paper, we have adopted a fixed mass flux of $\dot{M}_{\rm wind}=1 M_{\odot} \,\rm yr^{-1}$ over a timescale of $10\,\mathrm{Myr}$, which is consistent with  \citet{Veilleux_etal_2005ARA&A..43..769V} and \citet{Strickland_Heckman_2007ApJ...658..258S}.

\subsection{Ambient medium}
To investigate how the ambient medium affects the collimation and confinement of the wind, as well as how it alters the shape of the galactic bubble formed by the wind, we study two stationary profiles for the ambient medium. In all our simulations, the ambient medium is assumed to be in magnetohydrodynamic equilibrium and, in magnetized cases, a force-free magnetic field is set. Additionally, we ensure that the unperturbed, stratified ambient medium remains in hydrostatic equilibrium at all times.

We assume a structured ambient medium with a turbulent, warm disk and tenuous, hot halo shaped by the galactic rotation and the gravitational field according to \cite{Strickland-Stevens-2000}. The resulting effective gravitational potential $\Phi_{\rm all} = \Phi_{\rm ss} +\Phi_{\rm disk}$ from the stellar spheroid $\Phi_{\rm ss}$ with a King distribution profile reads
\begin{equation}\label{Eq:Potential_cluster}
    \Phi_{\rm ss}(r,z) = \frac{G M_{\rm ss}}{r_{0}}\left(\frac{\ln{\left[\left(R/R_{0}\right)+\sqrt{1+\left(R/R_{0}\right)^2}\right]}}{\left(R/R_{0}\right)}\right) \,,
\end{equation}
and galactic disk potential 
\citep{Miyamoto_Nagai1975PASJ...27..533M}  $\Phi_{\rm disk}$ has the following expression
\begin{equation}\label{Eq:Potential_disk}
    \Phi_{\rm disk}(r,z) = -\frac{G M_{disk}}{\sqrt{r^2+(r_{\rm disk}+\sqrt{z^2+z_{\rm disk}^2})^2}}\,,
\end{equation}
Here, $R=\sqrt{r^2+z^2}$ represents the spherical radial distance, $M_{\rm disk}$ is the disk mass, $M_{\rm ss}$ is the stellar spheroid mass, and $R_0$ is the core radius. The terms $r_{\rm disk}$ and $z_{\rm disk}$ denote the cylindrical radial and the vertical scale sizes for the disk, respectively. Following are the initial ambient medium density and pressure:

%
\begin{eqnarray}\label{Eq:density_pressure_all}
n&=&n_{\rm halo}+n_{\rm disk},\\
p&=&m_p\times\left(n_{\rm halo}\times c_{s,\rm halo}+n_{\rm disk}\times c_{\rm disk}\right),
\end{eqnarray}
where m$_{p}$ is the proton mass, $c_{s,\rm halo}$ and $c_{s,\rm disk}$ are the sound speeds of the halo and disk, respectively. $n_{\rm halo}$ and $n_{\rm disk}$ are respectively the halo and disk number density, and they can be expressed as follows:
\begin{gather}\label{Eq:Density_profile_hallo}
   n_{\rm halo}(r,z)=n_{\rm halo,0}\; \\ \exp{\left(-\frac{\Phi_{\rm all}(r,z)-e_{\rm h}^2\Phi_{\rm all}(r,0)-\left(1-e_{\rm h}^2\right)\Phi_{\rm all}(0,0)}{c_{s,\rm halo}^2}\right)}\,,\nonumber 
\end{gather}

\begin{gather}\label{Eq:Density_profile_disk}
   n_{\rm disk}(r,z)=n_{\rm disk,0} \\ \exp{\left(-\frac{\Phi_{\rm all}(r,z)-e_{\rm disk}^2\Phi_{\rm all}(r,0)-\left(1-e_{\rm disk}^2\right)\Phi_{\rm all}(0,0)}{c_{s,\rm disk}^2+\sigma^2_{\rm disk}}\right)}\,,\nonumber
\end{gather}
where $\sigma_{\rm disk}$ is the turbulent dispersion velocity. Finally, $e_{\rm h}$ and $e_{\rm disk}$ quantify the fraction of rotational support of the ISM for the halo and disk, respectively.


We adopt a supersonic turbulent velocity of $\sigma_{\rm disk}=75$ km/s, derived from \citet{Cooper_etal_2008ApJ...674..157C}, to ensure a realistic disk thickness. This choice of turbulence aims to maintain a reasonable disk thickness and prevent the occurrence of artificially high temperatures, as observed in previous studies such as those by \cite{Strickland-Stevens-2000}. The supersonic turbulence can arise from various sources, including spatial nonuniformities in star formation and thus of the galactic wind, radiation fields and supernovae, as well as hydrodynamic and magnetohydrodynamic instabilities such as the magnetorotational instability \citep[e.g.][]{Wada_Norman_2007ApJ...660..276W}.
The other parameters of the disk are chosen based on the paper by \citet{Strickland-Stevens-2000}, with a disk mass of $M_{\rm disk}=2 \times 10^9 M_{\odot}$, a disk radial scale of $r_{\rm disk}=222$ pc, and a disk vertical scale of $z_{\rm disk}=75$ pc. The rotation coefficient is set to $e_{\rm disk}=0.9$ for a thick disk and $e_{\rm disk}=0.95$ for a thin disk, respectively. These values are derived   by \citet{Strickland-Stevens-2000} to approximately reproduce M82's rotation curve.

The thickness of the disk in the case of a thin disk depends on the distance from the galactic center. In the inner region extending up to a distance of 1 kpc from the galactic center, the galactic disk exhibits a progressive increase in height. Specifically, at a cylindrical radial distance of $r=0.5$ kpc, the disk exhibits an approximate height of $z=0.5$ pc. It reaches its maximum height of $z=1.2$ kpc at a cylindrical radial distance of $r=1$ kpc.
In the case of the  thick disk scenario, at a cylindrical radial distance of $r=0.5$ kpc, the estimated height of the disk is approximately $z=0.8$ kpc. The maximum height of $z=1.2$ kpc is achieved at a larger cylindrical radial distance of $r=1.5$ kpc.

Regarding the mean values, we follow \citet{Melioli_2013} with the mean number density $n_{\rm disk,0}=100$ cm$^{-3}$, mean temperature $T_{s,\rm disk}=6.5 \times 10^{4}$ K (resulting in a mean sound speed of $c_{s,\rm disk}=300$ km/s).

The galactic halo is characterized by the following parameters \citep{Strickland-Stevens-2000, Strickland_2002,Cooper_etal_2009ApJ...703..330C, Melioli_2013}: stellar spheroid mass $M_{\rm ss}=6 \times 10^9 M_{\odot}$ for thin disk model and $M_{\rm ss}=3.6 \times 10^9 M_{\odot}$ for thick disk with a core radius of $R_{0}=350$ pc. It is assumed that the halo is not supported by rotation, and therefore the rotation coefficient is set to $e_{\rm halo}=0$. Concerning the mean number density, we use $n_{\rm halo,0}=0.2$ cm$^{-3}$, and for the mean temperature, we adopt $T_{s,\rm halo}=6.5 \times 10^{6}$ K, following \citet{Melioli_2013}.

When only the galactic halo is included in the model, the disk number density $n_{\rm disk,0}$ is set to zero in equation~\ref{Eq:density_pressure_all}.
These scenarios are particularly valuable for comparing our simulations with analytic models of wind bubble evolution, where the external medium is often assumed to be uniform \citep{weaver1977,koo1992}.
The settings used here follow the methodology employed in \citet{Strickland-Stevens-2000}. 
All the parameters characterizing the galactic disk and halo medium are summarized in Table~\ref{tab:table_ism_parameters}.



We consider three types of magnetic field configurations for the external medium: a uniform magnetic field parallel to the galactic axis, a dipolar magnetic field, and a mixed configuration combining both uniform and dipolar magnetic fields.
For the uniform cases, we are investigating two different magnetic field strengths: $10^{-5}$ Gauss and $10^{-4}$ Gauss. We selected these two cases because, based on our investigation, weak magnetic fields on the order of $10^{-6}$ Gauss and lower have a negligible influence on the evolution of the bubble during the initial $10 \mathrm{Myr}$. In fact, with such a magnetic field strength, the critical radius at which the magnetic field would start to significantly affect the bubble's evolution is $R_{\rm B} > 50 \mathrm{Mpc}$, as indicated by equation~\ref{Eq:Critical_radius_B} in section~\ref{Sec:The physics of the wind bubble}. The influence of these magnetic fields on the bubble begins after at least $25 \mathrm{Myr}$. On the other hand, when the uniform magnetic field exceeds $10^{-4}$ Gauss, the Alfvén speed in the halo approaches the speed of light at large distances from the galactic center.

For the dipolar cases, we consider magnetic field strengths of $B_{0}=10^{-5}$ Gauss, $B_{0}=10^{-4}$ Gauss, $B_{0}=10^{-2}$ Gauss, and $B_{0}=1$ Gauss. We choose to show the cases with this range of the magnetic strength, to demonstrate the influence of the magnetic field on the galactic bubble and to ensure that the Alfvén speed remains within a physical range. Indeed, the cases with a dipole with magnetic strength less than $B_{0}=10^{-6}$ Gauss, the magnetic field does not influence the shocks, and the bubble expands as in the non-magnetized case, even over a long evolution time, since with a dipole magnetic field strength drops with the radial distance as $1/R^3$ thus faster than the decrease of the wind ram pressure $1/R^2$. On the other side, the dipolar magnetic field with strength larger than $B_{0}=1$ Gauss, the Alfvén speed becomes of order of speed of light in the galactic free wind zone. The magnetic field observations in M82 conducted by \citet{lopezrodriguez2021} reveal a magnetic field strength on the order of mGauss within the bulk of the galaxy, which decreases to $\mu$Gauss at kpc scales. In the wind region, the stretched magnetic field exhibits a decrease in strength over a short distance. For instance, in the case of a dipole with $10^{-2}$ Gauss  we investigate, the strength of the stretched magnetic field drops to the mGauss level at a scale of approximately 500 parsecs. This is remains consistent with previously observed values \citep{Thompson_etal_2006,Adebahr_etal_2017A&A...608A..29A,lopezrodriguez2021}.  

Lastly, our investigation focused on a mixed configuration featuring an external medium characterized by a dipole magnetic field strength of $10^{-2}$ Gauss and a uniform magnetic field strength of $10^{-6}$ Gauss. It is worth noting that we deliberately chose to explore a scenario with a dominant dipolar magnetic field component. This choice was motivated by the fact that, within the free galactic wind zone, the magnetic field undergoes stretching, causing its average strength to drop to the order of $\mu$Gauss. Furthermore, we opted for a weak uniform magnetic field component to ensure that its impact on the dynamics of the bubble remains subordinate to that of the dipolar component within a 1 kpc zone. Based on observational data presented by \citet{lopezrodriguez2021}, the strength of the magnetic field is primarily measured in the free galactic wind region (1-2 kpc from the core), where it is estimated to be on the order of 1–100 $\mu$Gauss. However, as demonstrated in the simulations conducted in this paper, the galactic wind stretches the magnetic field in this region, resulting in a weaker strength than what would be expected from a dipolar field.

The details of all the configurations studied in this paper are summarized in Table~\ref{tab:init_cond}.

\section{Dynamical evolution}\label{Sec:Dynamical evolution}

\begin{table*}
	\centering
	\caption{
	Galactic medium parameters according to \citet{Strickland-Stevens-2000,Cooper_etal_2008ApJ...674..157C,Melioli_2013}.}
	\begin{tabular}{lcc}
	\hline
	Parameters &  Thick-disk models & Thin-disk models\\
	\hline
    \hline
	stellar spheroid (galactic core) mass    M$_{rm ss}$ (M$_{\odot}$)      &  $3.6\times 10^{8}$  &  $6\times 10^{8}$ \\
    core radius R$_{0}$ (pc) & 350 & 350\\
    galactic halo mean density (cm$^{-3}$) & 0.2 & 0.2\\
    galactic halo mean temperature (K) & $6.7\times 10^6$&$6.7\times 10^6$ \\
    galactic disk mass    M$_{\rm  disk}$ (M$_{\odot}$)      & $6\times 10^{9}$  & $6\times 10^{9}$  \\
    galactic disk radius scale r$_{\rm disk}$ (pc) & 222 & 222\\
    galactic disk vertical scale r$_{\rm disk}$ (pc) & 75 & 75\\
    galactic  disk mean density $n_{\rm disk, 0}$ (cm$^{-3}$)& 100 & 100\\
    galactic disk mean temperature (K) & $6.7\times 10^4$&$6.7\times 10^4$ \\
    galactic disk turbulent speed (km/s) &75 &75\\
    galactic disk rotation coefficient $e_{\rm disk}$ &0.95&0.9\\
	\hline\\ 
	\end{tabular}
\label{tab:table_ism_parameters}
\end{table*}
The simulations are performed using the Message Passing Interface-Adaptive Mesh Refinement Versatile Advection Code (MPI-AMRVAC; \citealt{Keppens_etal_2021}). We utilize the MPI-AMRVAC magnetohydrodynamics module to resolve the mass, momentum, energy, and induction equations. The effect of optically thin radiative cooling is included as a source term in the energy equation \citep{VanMarleKeppens_2011CF.....42...44V}. The following equations are solved:
 \begin{eqnarray}\label{Eq:S_EQ_MHD_conserved}
 \frac{\partial \rho}{\partial t}~+~\vec{\nabla}\cdot(\rho\vec{v})~ = 0 \,, \label{eq:euler:mass}\\
\frac{\partial \rho\vec{v}}{\partial t}~+~\vec{\nabla}\cdot(\rho\vec{v}\cdot\vec{v}+\vec{B}\cdot\vec{B})~+\nabla \left(p+\frac{B^2}{2}\right)~= 0 \,,\label{eq:euler:momentum} \\
\frac{\partial e}{\partial t}~+~\vec{\nabla}\cdot\left( \left(e+p+\frac{B^2}{2}\right)\vec{v}-\vec{B}\vec{B}\cdot\vec{v}\right)~ = -\left(\frac{\rho}{m_{h}}\right)^2\Lambda(T)\,, \label{eq:euler:energy}\\
\frac{\partial \vec{B}}{\partial t}+\vec{\nabla}\times\left(\vec{v} \times \vec{B} \right)\,=\,0\,.
            \label{eq:euler:induction}
 \end{eqnarray}
In the above equations, $\rho$ represents the mass density, $\vec{v}$ is the velocity, $p$ is the thermal pressure, and $m_{h}$ is the mass of hydrogen. The total energy density of the gas is denoted by $e= B^2/2 + p/(\gamma-1) + \rho v^2/2$, where the constant polytropic index $\gamma$ is set to $\gamma = 5/3$. 
 $\vec{B}$ represents the magnetic field.
The divergence of the magnetic field is maintained close to zero using the divergence cleaning method \citep{Dedner_2002JCoPh.175..645D}.
In the energy equation, $\Lambda(T)$ represents the optically thin radiative cooling function, which is dependent on the local temperature $T$ and the chemical composition.
 In all the simulations conducted, we consistently adopt an atomic cooling curve that is calibrated for solar-metallicity. Specifically, we utilize the cooling function $\Lambda(T)$ as presented in the work of \citet{Schure_etal_2009A&A...508..751S} and also as used in \citep{Yu_etal_2020MNRAS.492.3179Y}.
It is important to note that the iron abundance in M82 does not precisely match that of the sun \citep{Origlia_2004ApJ...606..862O}. Nonetheless, in the scenario under investigation, the primary cooling occurs in the forward shock, which is situated between the galactic wind and the galactic disk. In this region, the temperature is around $T\approx 10^5$ K, rendering the differences in ion-abundance inconsequential. 

Throughout the simulations, we maintain a floor temperature of 50 K, where we neglect the effect of photoionization. Furthermore, cooling is not applied to the non-shocked external medium, and the temperature of the non-shocked galactic halo and galactic disk remains fixed during the simulations. We assume that the non-shocked galactic disk and halo are in thermal equilibrium, balancing radiative cooling with the heating from stellar radiation, cosmic-ray pressure and waves dissipation \cite[e.g.][]{Girichidis_etal_2018MNRAS.479.3042G}. This equilibrium prevents the hot galactic halo and disk from expanding outward. It is important to note that the imposed floor temperature is not reached in the simulations investigated in this paper, and is set to avoid nonphysical cooling and for numerical stability reasons as well.

The simulations are conducted on a 2D spherical grid with a range of $\left[130,13000\right]$ pc in the radial direction and $\left[0,\pi\right]$ in the polar direction. A logarithmic grid is used in the radial direction. At the first level of refinement, a resolution of $(160\times96)$ cells is employed, corresponding to a cell size of $\Delta \theta\approx 1.8^\circ$ in the polar direction. In the radial direction, the first cell at the inner boundary has a size of $\Delta R\approx 2.279\, \mathrm{pc}$, increasing logarithmically toward the outer boundary.

Due to the large-scale ratio between the shock scales and the total simulation box investigated in this project, Adaptive Mesh Refinement (AMR) is crucial for adequately resolving the shocks and accurately treating the cooling. In the performed simulations, the grid is allowed to be refined up to 7 additional levels, with a doubling of resolution at each new level of refinement, resulting in a cell size of $\Delta \theta \approx 0.028^\circ$ in the angular direction. The first cell near the inner boundary has a radial size of $\Delta R\approx 0.0356 \,\mathrm{pc}$.
The refinement and coarsening are accomplished using Lohner's error estimation method \citep{Lohner_2011IJNMF..67.2184L} to assess variations in density and temperature. The variation in temperature enables AMR to resolve contact discontinuities, compression waves, and shock waves. Additionally, using the temperature variation is crucial for accurately resolving regions characterized by strong compression and/or intense cooling.


Concerning the boundary conditions, the polar axis is treated as an axial boundary. At the radial inner boundary, a fixed stellar mass-loss rate $\dot{M}_{\rm wind}=1\,M_{\odot} \, \rm yr^{-1}$, speed $v_{\rm wind} =1000 \, \rm km \, s^{-1}$ \cite{Melioli_2013} and Mach number of $\mathcal{M}_{\rm wind} = 1.01$ are imposed during the $10$~Myr simulation time. The radial outer boundaries are treated as open boundaries, using a Neumann boundary condition with vanishing derivatives for all quantities. For the spherical radial magnetic field component, free divergence is used at the boundaries.


The fluid equations (\ref{eq:euler:mass}-\ref{eq:euler:induction}) are solved using the Total Variation Diminishing (TVD) method \citep{Harten_1983JCoPh..49..357H}, specifically the Harten-Lax-van Leer-Contact (HLLC) method \citep{Toro_etal_1994ShWav...4...25T}, which allows for the resolution of shocks and contact discontinuities. This is combined with the Koren limiter \citep{Koren_bfe87ab1e06744e7b6e2163728ba7ac4}, a third-order asymmetric TVD limiter.

For simulations with strong magnetic fields, such as the dipole configuration with $B_0=1$ Gauss, the numerical scheme is switched to TVDLF (Total Variation Diminishing Lax-Friedrichs) \citep{Toth_1996JCoPh.128...82T} and the minmod limiter \citep{Roe_1986AnRFM..18..337R} in the vicinity of the polar axis to ensure accurate results. This switch reduces the numerical instability that could develop near the polar axis when employing a more accurate numerical scheme that combines HLLC and the Koren limiter.

 \begin{table*}[!h]
     \centering
     \begin{tabular}{|l|c|c|c|c|}
     \hline
         Cases &  external medium  profile &Magnetic field structure & B$_{\rm z}$ [Gauss]& B$_{\rm 0}$ [Gauss]\\
         \hline
         A   & halo and thick disk  &---&   0.0&0.0\\
          B   & halo and thin disk  & ---&  0.0&0.0\\        
         \hline
         C   & halo   &  uniform &$10^{-5}$ &0.0\\    
         D   & halo   &  uniform& $10^{-4}$& 0.0\\          
         \hline     
         E   & halo   &  dipole& 0.0&$10^{-5}$\\    
         F   & halo   &  dipole& 0.0&$10^{-4}$\\  
         G   & halo   &  dipole& 0.0&$10^{-2}$\\
         H   & halo   &  dipole& 0.0&1\\   \hline
         I   & halo and thick disk  &  dipole& 0.0&$10^{-4}$\\   
         J   & halo and thin disk  &  dipole& 0.0&$10^{-4}$\\       
         K  & halo and thick disk  & dipole& 0.0&$10^{-2}$\\       
         L   & halo and thin disk  &  dipole& 0.0&$10^{-2}$\\        
        M   & halo and thick disk   & dipole& 0.0&$1$\\   
        N   & halo and thin disk    & dipole& 0.0&$1$\\    
        P   & halo and thick disk  &  dipole+uniform B$_{\rm z}$& $10^{-5}$&$1$\\   
        Q   & halo and thin disk  &  dipole+uniform B$_{\rm z}$& $10^{-5}$&$1$\\            
         \hline 
         \end{tabular}
         \caption{Parameters describing the ambient medium in starburst galaxies across various cases studied in this paper are presented. For each case, the external medium's density and pressure profile, magnetic field structure, and strength are provided. All cases share the same properties for the galactic wind: a mass-loss rate of $\dot{M}_{\rm w} = 1\,\mathrm{M\odot/yr}$, a speed of $v_{\rm w} = 2000\,\mathrm{km/s}$, a Mach number of $\mathcal{M} = 20$, and a duration of $t = 10$ Myr. }
         \label{tab:init_cond}
\end{table*}

\section{Results}
In this section we report the results of a set of simulations (see Table \ref{tab:init_cond}) performed in different conditions of external medium properties, including both disk and halo, and different assumptions in the large-scale magnetic field. In this context, we particularly focus on the geometrical properties of the resulting wind bubbles, as well as the strength of the shocks and, consequently, their capability of accelerating particles via diffusive shock acceleration.


\subsection{Unmagnetized halo-disk model}
\begin{figure}[!h]
\includegraphics[width=.35\textwidth,angle=-90]{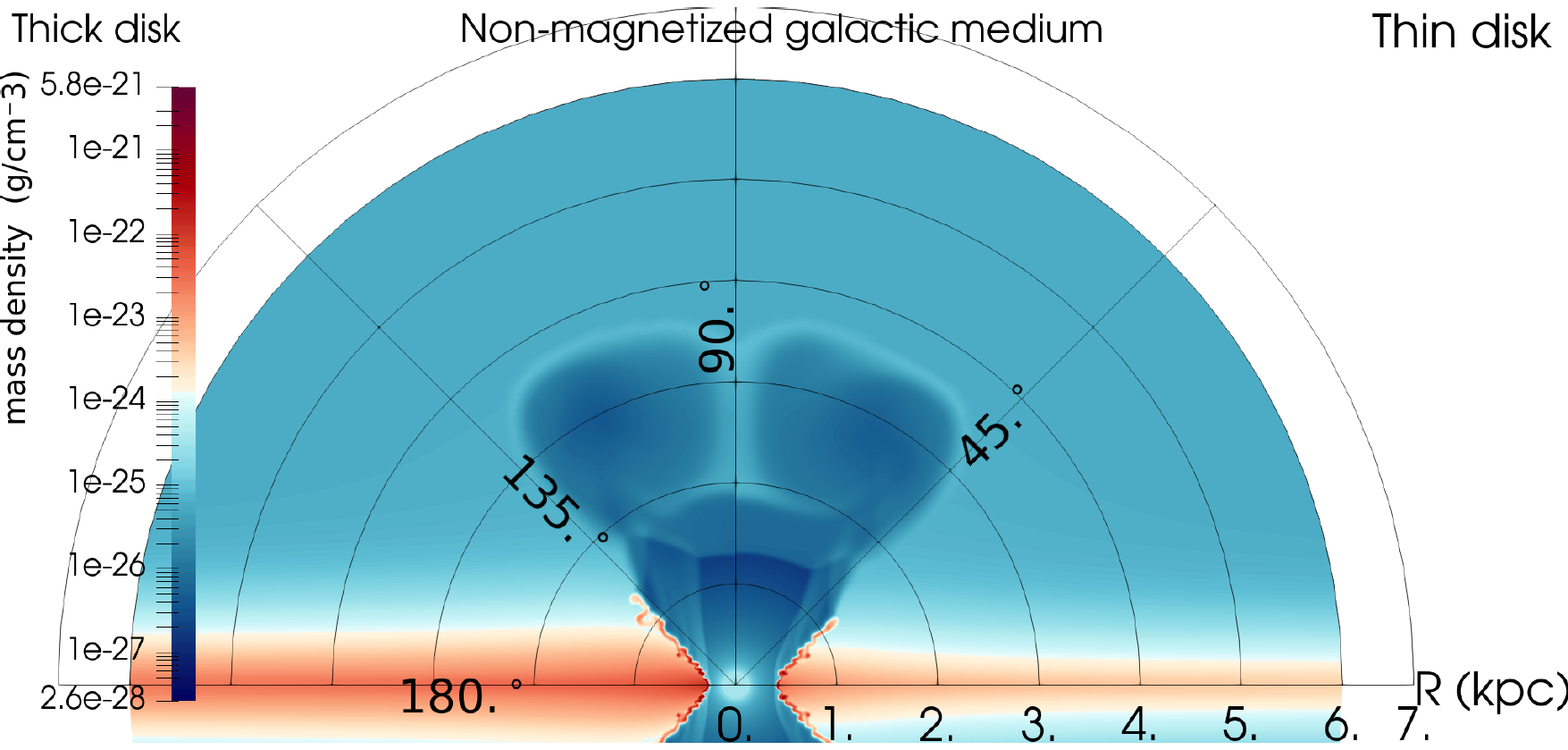}

\caption{Density contour for the HD case at t=10 Myr with slower galactic wind $v=1000 \mathrm{km/s}$. The left panel shows the thin disk and the right panel shows the thick disk, both with a galactic wind of 
$\dot{M}$= 1 M$_{\odot}$/yr.}
\label{fig:HDcooper}
 \end{figure}

In starburst galaxies such as M82, the surrounding galactic medium is distributed under the gravitational potential resulting from the stars and gas distribution and rotation profile of the galaxy. The presence of the galactic disk can influence the evolution of the wind bubble, as demonstrated by previous studies by authors such as ~\citep[e.g.][]{Strickland-Stevens-2000,Strickland_2002,Cooper_etal_2008ApJ...674..157C,Melioli_2013}. 


We first examine the non-magnetized cases referred to as models A and B in Table \ref{tab:init_cond}. Model A represents a scenario with a thick galactic disk, whereas model B corresponds to a thin disk configuration.
Figure~(\ref{fig:HDcooper}) illustrates the influence of these two distinct galactic disk models: the thick disk is shown on the left, and the thin disk on the right. The figure illustrates the characteristics of the outflow at an age of 10 Myr.

A prominent shock near the equator is induced by the strong interaction between the galactic wind and the galactic disk. While maintaining the same inclination with the galactic axis, this shock expands in higher altitude beyond the disk toward the galactic halo. The shock produces a dense shell that confines the galactic outflow along the axis, leading to a collimated outflow. The half-opening angle is approximately $\theta \approx 45^{\circ}$  for the thin disk model and $\theta \approx 40^{\circ}$ for the thick disk model.

In the free galactic wind region, the wind is accelerated thermally to reach a speed of $v_{\rm w}\,\approx 1800 \mathrm{km/s}$. This acceleration occurs within $0.5 \mathrm{kpc}$. As a result, within the wind funnel, the terminal wind speed is the same in both the thin and thick disk cases. Consequently, the termination shock has the same speed close to the galactic axis, even if the lateral extent is different.
In the upper region of the forward shock front, the confined outflow interacts with the hot, rarefied halo. This interaction results in a less pronounced forward shock, as a fraction of the shocked gas from the halo moves toward the galactic axis, thereby increasing the mass loading of the outflow. As a result, a distinctive “hat” structure is formed near the galactic axis. In the thin disk and thick disk cases, the bubble reaches the same heights of  $3.63 \mathrm{kpc}$.

\begin{figure*}[h!]
\centering
\includegraphics[width=.58\textwidth,angle=-90]{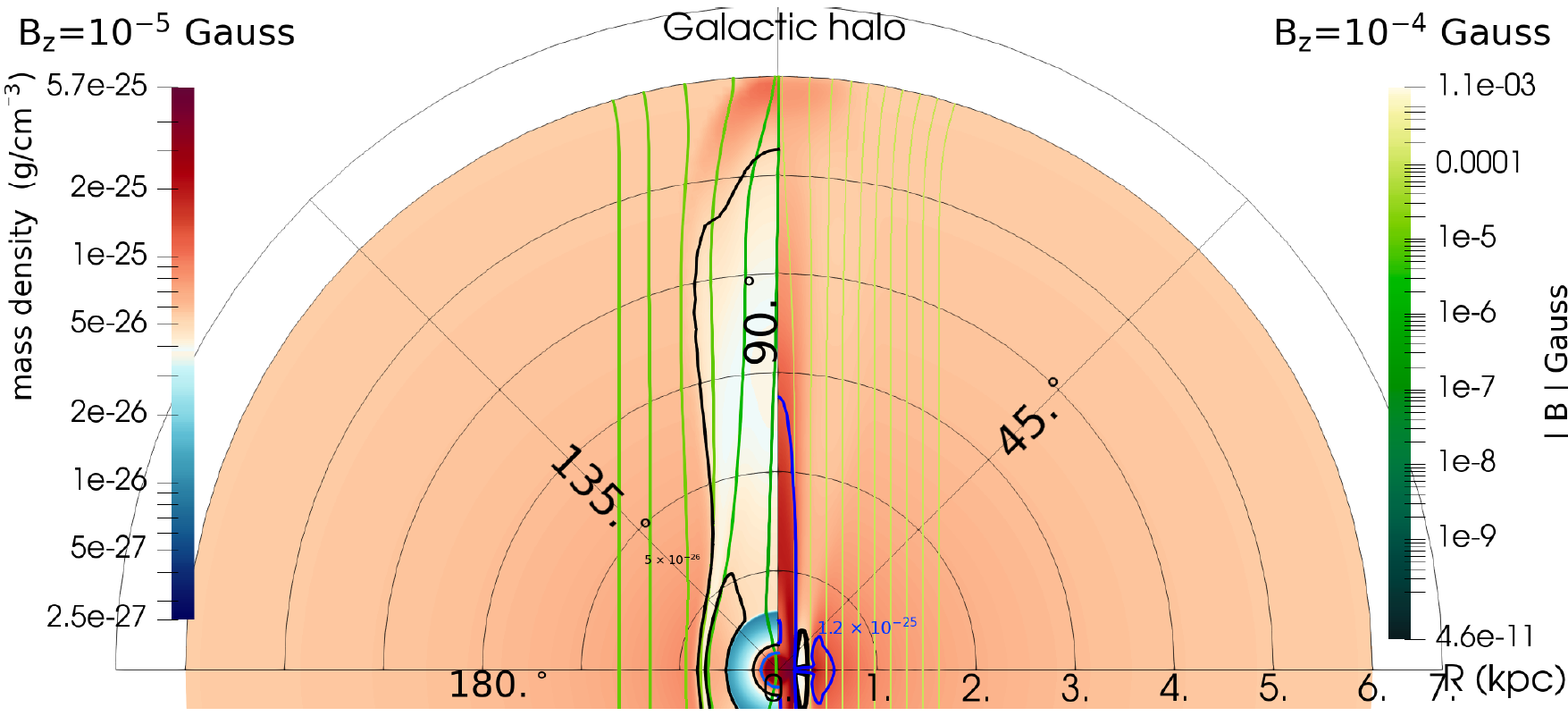}
\caption{Density contour of an expanding galactic bubble in a halo with a uniform magnetic field at $t=10$ Myr, accompanied by a mass loss from the galactic wind, denoted as $\dot{M}= 1 \, \mathrm{M_{\odot}/yr}$. The external medium has a magnetization of $B_{\rm z} =10^{-5}$ Gauss (left) and $B_{\rm z} =10^{-4}$ Gauss (right). The density isocontours are represented by blue and black lines, corresponding to their respective density isocontour values of $5\times 10^{-26}$ and $1.2\times 10^{-25}$ respectively.
}
\label{fig:uniform1}
\end{figure*}

At the age of 10 Myr, near the equatorial plane, both the termination shock and forward shock are quasi-stationary and exhibit high intensity with a compression rate approximately $10$. The shocks occur at a distance of 0.4 kpc in the case of the thin galactic disk, and at a distance of 0.25 kpc in the case of the thick galactic disk.

However, as we move toward the galactic axis, the strength of these shocks gradually diminishes. The termination shock has a compression rate of approximately $\mathcal{R}_{\rm ts}=\rho_{\rm ts, post}/\rho_{\rm ts, pre}\approx 4$ and it expands at a speed $v_{\rm ts}\approx 100 \, \mathrm{km \, s^{-1}}$, with the associated Mach number of $\mathcal{M}_{\rm ts}=v_{\rm ts}/c_{\rm s, w}\approx 50$ (where $c_{\rm s, w}$ is the sound speed of the galactic wind). The forward shock has a compression rate of $\mathcal{R}_{\rm fs}=\rho_{\rm fs, post}/\rho_{\rm fs, pre}\approx 1.8$ and it expands at a speed on the order of $v_{\rm fs}\approx 300$ km/s with a Mach number of $\mathcal{M}_{\rm fs}=v_{\rm fs}/c_{\rm s, h}\approx 1.5$ ($c_{\rm s, h}$ is the sound speed of the non-shocked halo).

It is important to note that in both the thick and thin disk scenarios, the size of the bubble at 10 Myr remains the same. However, the structure within the bubble is different, with a more extended region of shocked-reflected wind in the vicinity of the galactic disk. Indeed, the shocked-reflected wind restricts the free wind within an angle of 30$\degree$ in the thin disk scenario and 6$\degree$ in the thick disk scenario (Figure~\ref{fig:HDcooper}).

\subsection{Magnetized halo model without disk}

When a galactic bubble expands in the galactic halo (without a galactic disk), its deviation from a spherical shape depends mainly on the strength and configuration of the large-scale magnetic field. Indeed, instabilities could also contribute to the deformation of the shock's surface, which depends on the strength of the shocks. Here, we consider two geometrical configurations and various strengths of the magnetic field: a uniform magnetic field aligned with the galactic axis and a dipole.

\subsubsection{Uniform magnetic field cases}
In Fig.~\ref{fig:uniform1}, the effect on the galactic bubble of increasing the magnetic field strength from $B_{\rm z} = 10^{-5}$ Gauss to $B_{\rm z} = 10^{-4}$ Gauss (models C and D, respectively) is shown.

\begin{figure*}[h!]
\centering
\includegraphics[width=0.58\textwidth,angle=-90]{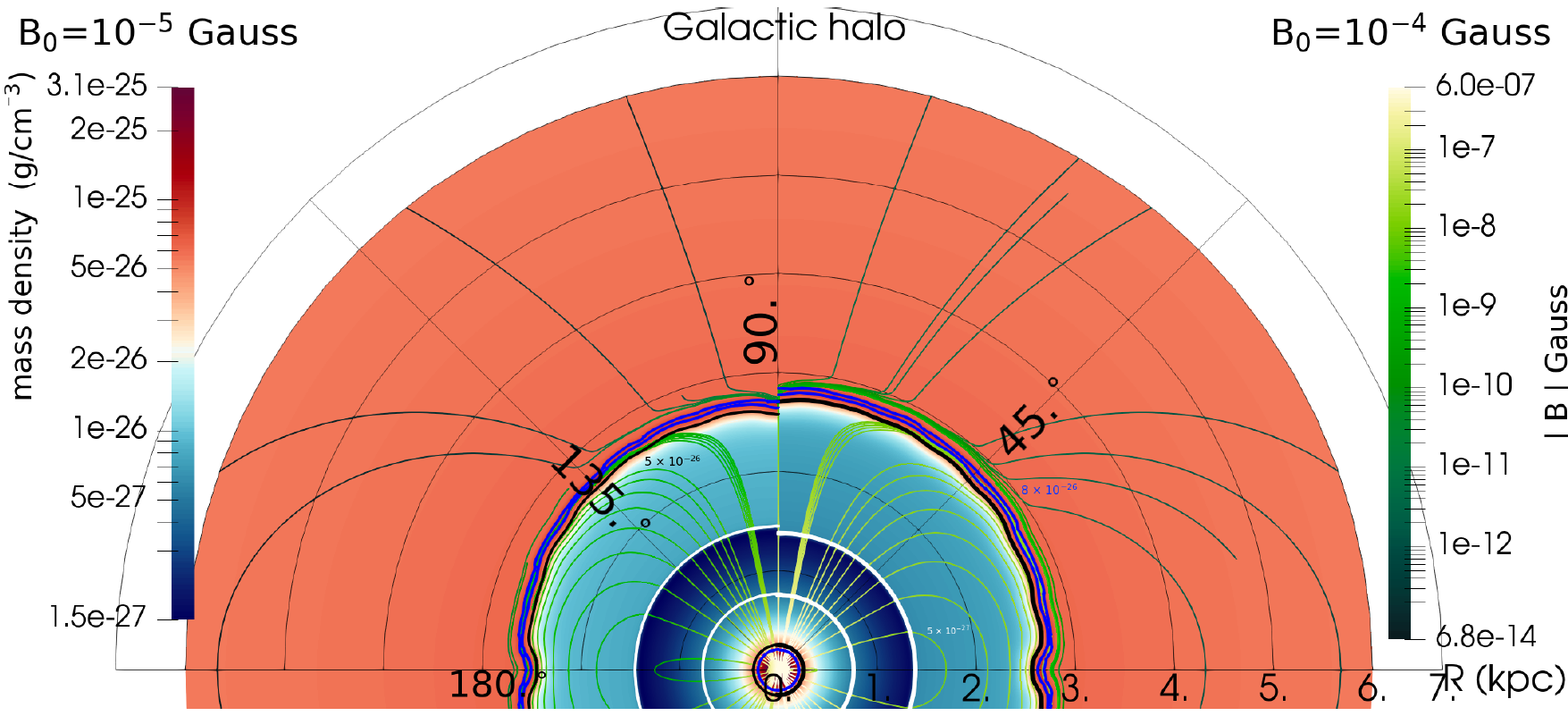}
\includegraphics[width=.58\textwidth,angle=-90]{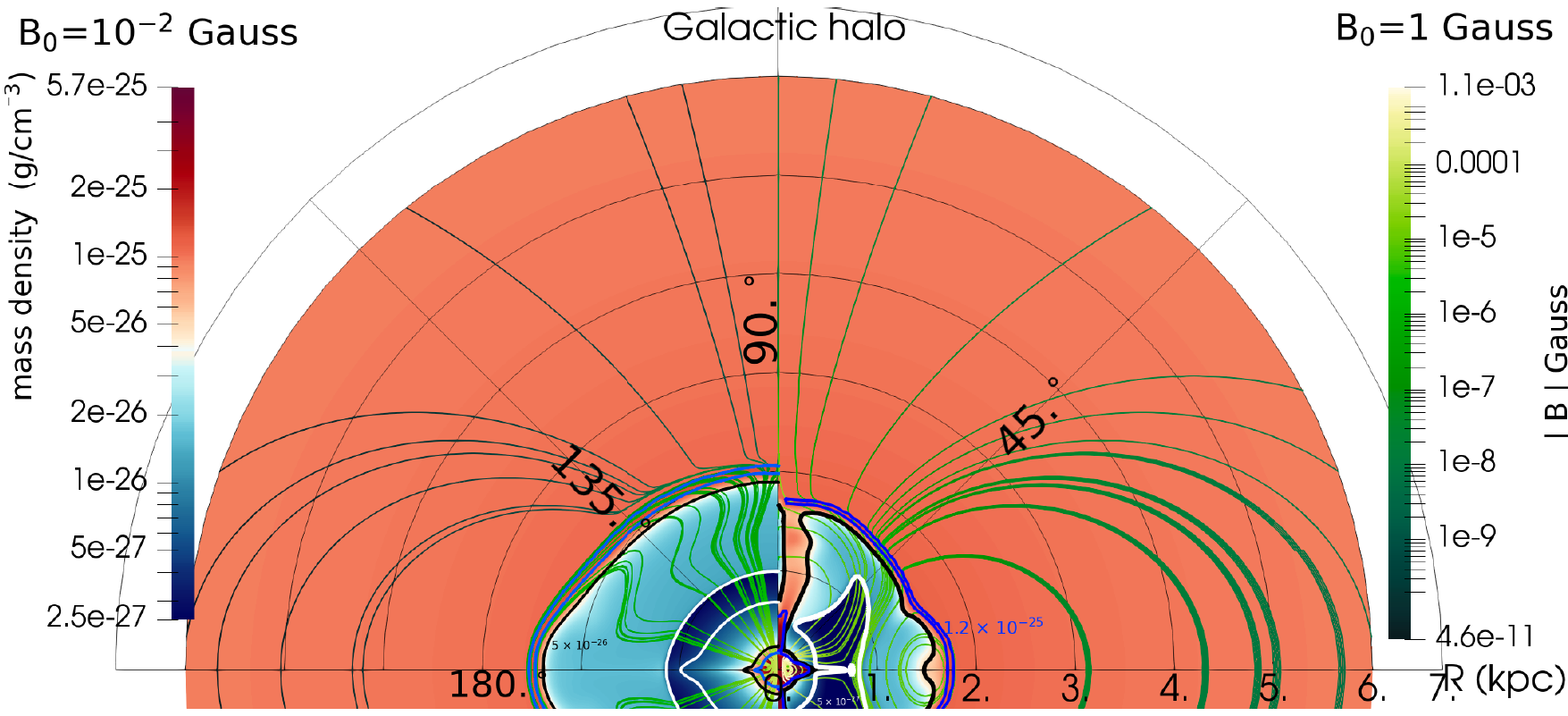}
\caption{Density contour for an expanding galactic bubble in a magnetized halo with a dipolar magnetic field at t=10 Myr. The galactic wind has a mass flux of $\dot{M}$= 1 M$_{\odot}$/yr. Top panel: $B_0 =10^{-4}$ Gauss (left) and $B_0 =10^{-5}$ Gauss (right). Bottom panel: dipole $B_0 =10^{-2}$ Gauss (left) and $B_0 =1$ Gauss (right). The isocontour of the density is represented by white, blue and black  lines, corresponding to their respective density isocontour values of $5\times 10^{-26}$, $5\times 10^{-26}$ and $1.2\times 10^{-25}$ respectively}
\label{fig:dipole1}
\end{figure*}

The expansion of the galactic wind induces pressure and tension forces on the halo matter and magnetic field lines. As a result, the compressed magnetic field lines exert forces that decelerate the expansion of the bubble in the direction perpendicular to the field lines. This leads to elongation along the galactic axis, which aligns with the direction of the magnetic field in this particular model.
Moreover, the strength of the forward shock diminishes gradually as we move from the polar axis. Indeed, at the axis the shock normal is parallel to the local magnetic field, and toward the equatorial plane, the shock normal becomes perpendicular to the magnetic field lines.

%
%

For a weak magnetic field with a strength of $B_{\rm z}=10^{-5} \mathrm{Gauss}$ (model C), the termination and the forward shocks exhibit distinct characteristics as they move from the galactic axis to the galactic equatorial plane.
The termination shock remains spherical with 
$\mathcal{R}_{\rm ts} = \rho_{\rm ts, post}/\rho_{\rm ts, pre}\approx 5$
and  a speed on the order of $v_{\rm ts}\approx 3.4$ km/s.
On the other hand, the  forward shock is anisotropic. 
In fact, it is  more elongated along the galactic axis. 
In the vicinity of this axis, the forward shock wave is  weak  with a compression ratio 
$\mathcal{R}_{\rm fs}= \rho_{\rm fs, post}/\rho_{\rm fs, pre} \approx 1.6$ while it expands with a speed on the order of $v_{\rm fs}\approx 1000 \mathrm{km/s}$ with a fast magnetosonic number on the order of one. 
At the equatorial plane, the forward shock turns into a weak compression wave and expands slowly with a speed of the order of $v_{\rm fs}\approx 10 $ km/s.

%
%
In the case of a relatively strong and uniform magnetic field with a strength of $B_{\rm z}=10^{-4}$ Gauss (referred to as model D), the expanding bubble is confined along the polar axis. The termination shock is intense, with a compression rate of $\mathcal{R}_{\rm ts}\approx 6$. It remains quasi-spherical and quasi-stationary.

In this specific scenario, after the first 1 Myr of evolution, the forward wave propagating through the ambient medium begins to weaken. At this point, it transitions from a shock wave to a weak compression wave. Additionally, the shocked halo material near the galactic axis assumes a jet-like shape.

This change in behavior can be explained by estimating the wind ram pressure, given by $P_{\rm wind}= \rho v_{\rm wind}^2\approx 0.5 \times 10^{-10} \,\mathrm{dyn/cm^2}$, at a distance of 0.5 kpc. Conversely, at the equator, the magnetic field pressure is $P_{\rm B}= B^2/(8 \pi)\approx 8 \times 10^{-10} \, \mathrm{dyn/cm^2}$ for model D with $B_{\rm z}=10^{-4}$ Gauss, which is 16 times greater than the wind's ram pressure. However, in model C where $B_{\rm z}=10^{-5}$ Gauss, the magnetic pressure $P_{\rm B}= 8 \times 10^{-12}\,\mathrm{dyn/cm^2}$ is still less than the wind ram pressure at 0.5 kpc.

\begin{figure*}[h]
\includegraphics[width=.35\textwidth,angle=-90]{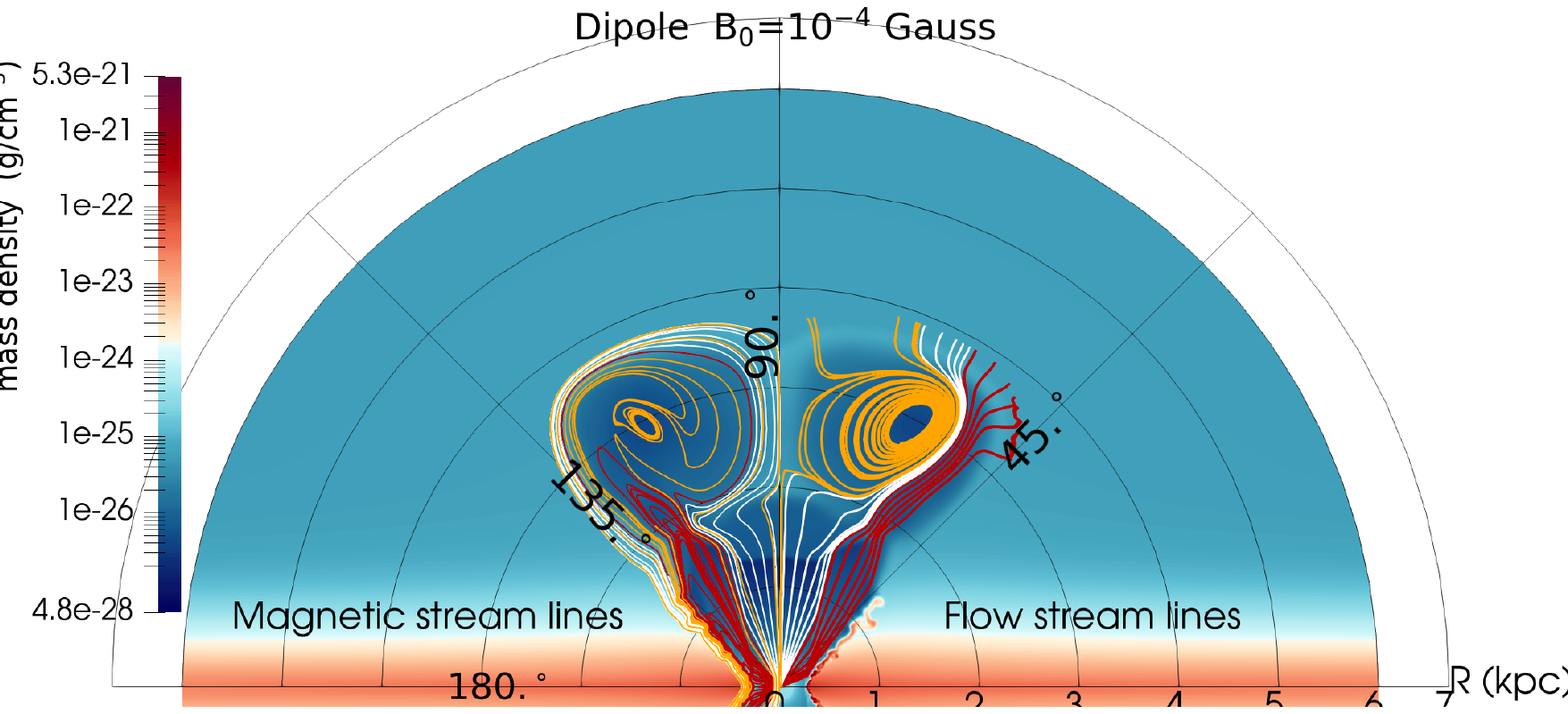}
\includegraphics[width=.35\textwidth,angle=-90]{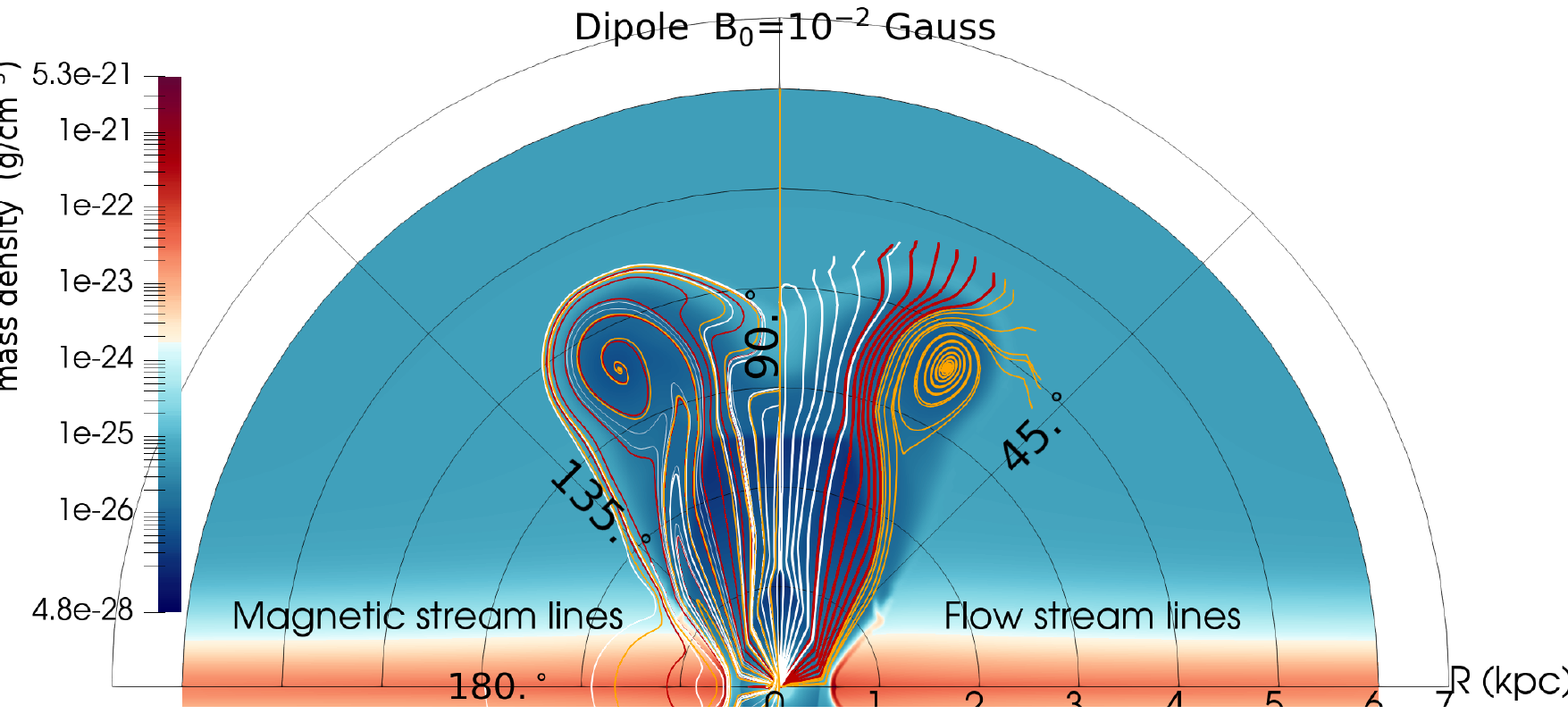}
\caption{Density contour, magnetic streamlines, and flow streamlines for an expanding galactic bubble with a large-scale magnetic field with a thick galactic disk at the left (B = $10^{-4}$ Gauss) and the right (B = $10^{-2}$ Gauss)  at t = 10 Myr. (Left) Density contour showing variations in density and the magnetic streamlines representing the direction and strength of the magnetic field, color-coded to indicate their emergence zone. (Right) Flow streamlines indicating the motion of matter, also color-coded to indicate their emergence zone.}
\label{fig:MHDB01e_4_stream_lines}
\end{figure*}
\subsubsection{Dipolar magnetic field cases}
Figure~\ref{fig:dipole1} demonstrates the impact of a dipolar magnetic field with central amplitudes ranging from $B_0=10^{-5}$ Gauss to $B_0=1$ Gauss on the expansion of the galactic bubble at an age of $t=10$ Myr.

For magnetic field strengths between $10^{-5}$ and $10^{-2}$ Gauss, the ram pressure of the free wind applies a force on the magnetic field lines, leading to their radial reshaping. This region is marked by significant stretching of the field lines, which results in a decrease in their strength.
In the area between the termination shock and the forward shock, the closed magnetic field lines impede the shell's expansion and cause deformation of the contact discontinuity surface.
At an age of 10 Myrs, a magnetized bubble with a magnetic field strength up to $10^{-4}$ Gauss exhibits a predominantly isotropic and strong termination shock that reaches a distance of 1.4 kpc in model (E) with $B_{0}=10^{-5}\,\mathrm{Gauss}$ and 1.3 kpc in model (F) with $B_{0}=10^{-4}\,\mathrm{Gauss}$. This minor difference is due to the magnetic field, which reduces the strength of the shocks and consequently slows down the expansion of the termination shock.
In the laboratory frame, the termination shock moves at a speed of $v_{\rm ts}\approx 100 \,{\rm km/s}$, and its compression ratio is $\mathcal{R_{\rm ts}}=\rho_{\rm ts, post}/\rho_{\rm ts, pre} \approx 1.3$. The forward shock is weak and has a speed on the order of $v_{\rm fs}\approx 125 \, {\rm km/s}$ and a fast magnetosonic number of $\mathcal{M_{\rm fs}}\approx 1$. It compresses the perpendicular components of the magnetic field, causing the angle between the magnetic field and the shock normal to increase (as shown in Figure \ref{fig:dipole1}). As a result, the forward shock becomes deformed by a weak instability that grows during the initial evolution phase before subsiding. However, as the distance from the source increases, the magnetic field of the dipole weakens and the magnetization at the forward shock decreases over time. Consequently, the strength of the forward shock remains relatively constant throughout the evolution.

In the case of a strong dipolar magnetic field with an amplitude $B_{0}=10^{-2}$ Gauss or $B_{0}=1$ Gauss, the magnetic field guides the non-shocked galactic wind. In this region, the Alfvén Mach number is approximately 2 (in the case $B_{0}=1$ Gauss). Consequently, the galactic wind exhibits a predominant flow pattern along 45 degrees from the galactic axis when the dipolar magnetic field is $B_{0}=1$ Gauss. The presence of the magnetic field significantly affects both the termination and forward shocks, leading to their deceleration and reshaping. The termination shock remains strong (as shown in Figure \ref{fig:dipole1}), with a compression ratio of approximately $\mathcal{R}_{\rm ts}\approx 4$. However, it is important to note that the compression ratio at the termination shock increases from the equator to the pole. This increase is due to the enhanced magnetic pressure and tension in the shocked wind and halo regions, which slow down the propagation of the termination shock. Additionally, in the case with $B_{0}=1$ Gauss, after 9 Myr, the termination shock outside the equatorial plane begins to propagate inward at a speed on the order of 10 km/s.
Regarding the forward shock, the compression ratio decreases from the equator, where $\mathcal{R}_{\rm fs} \approx 5.4$, to the pole, where $\mathcal{R}_{\rm fs} \approx 1.4$. This decrease indicates that it is weaker compared to models with a much weaker magnetic field. In the case with $B_{0}=1$ Gauss, after 9 Myr, the forward shock becomes quasi-stationary.

\begin{figure*}[h]
\includegraphics[width=.35\textwidth,angle=-90]{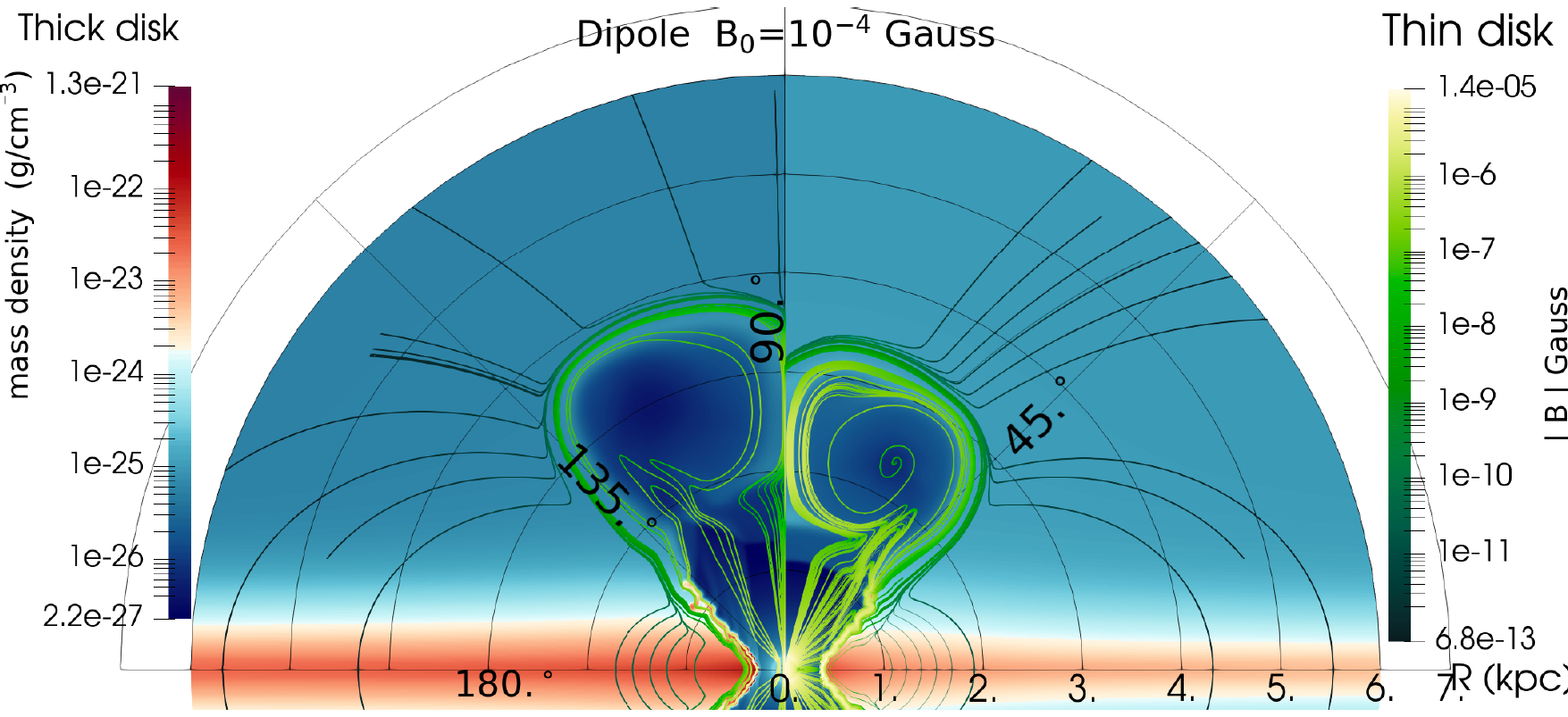}
\includegraphics[width=.35\textwidth,angle=-90]{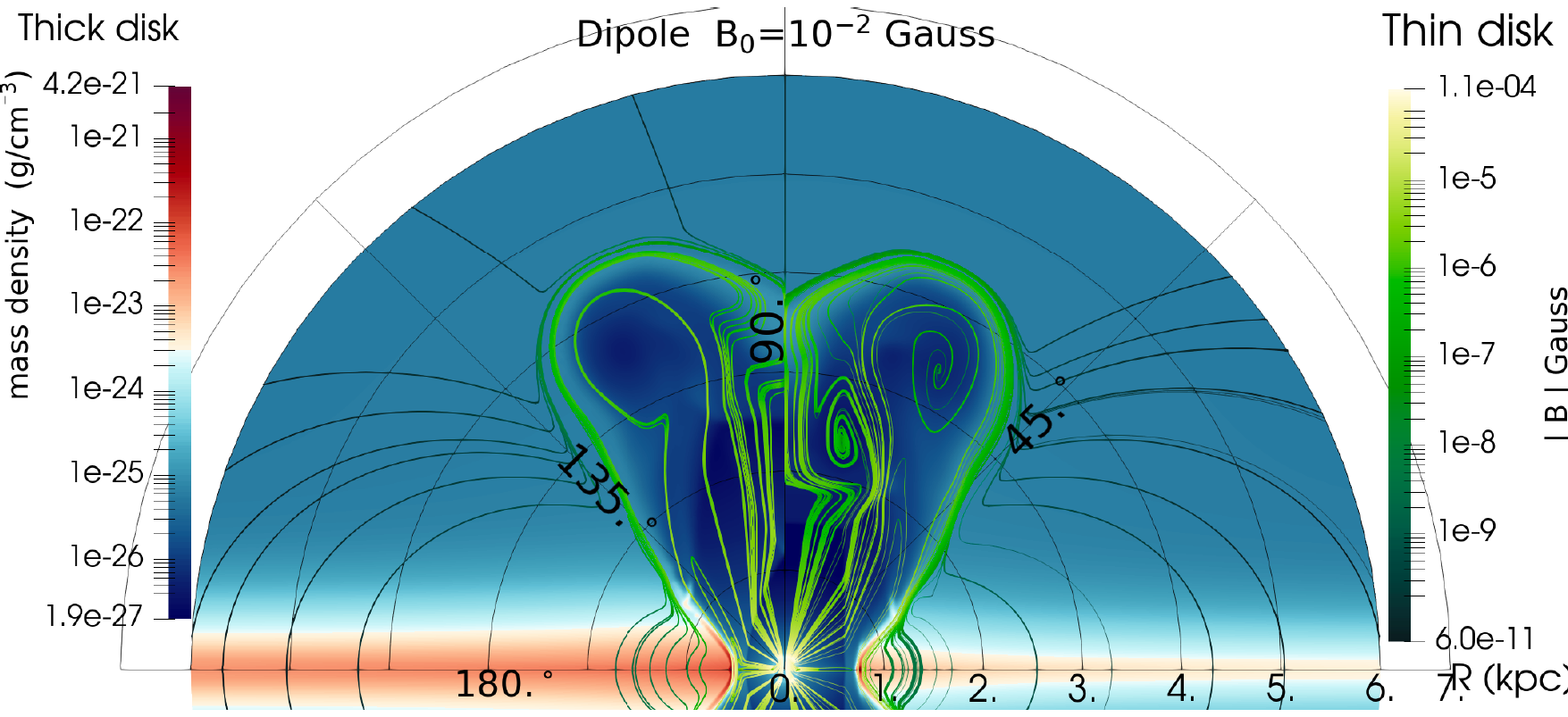}
\includegraphics[width=.35\textwidth,angle=-90]{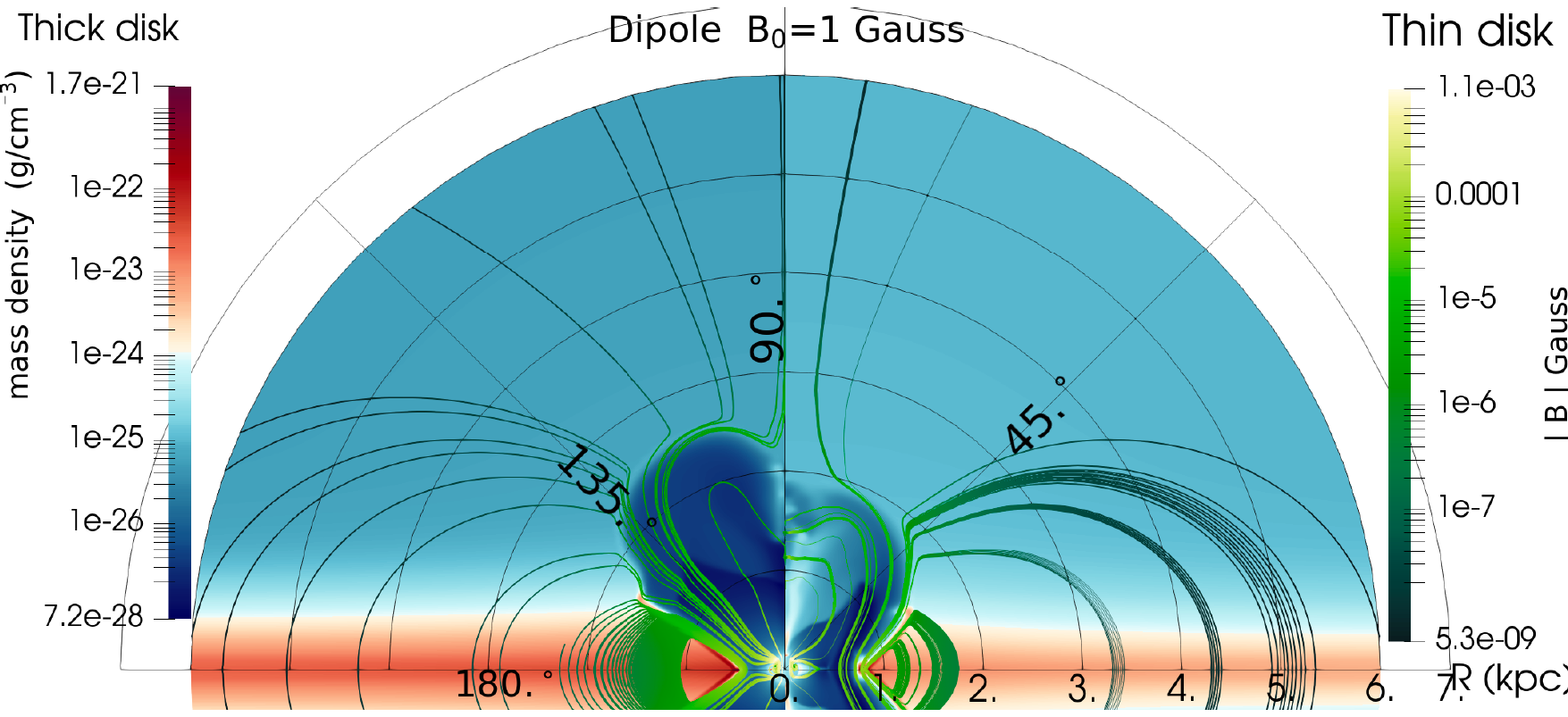}
\includegraphics[width=.35\textwidth,angle=-90]{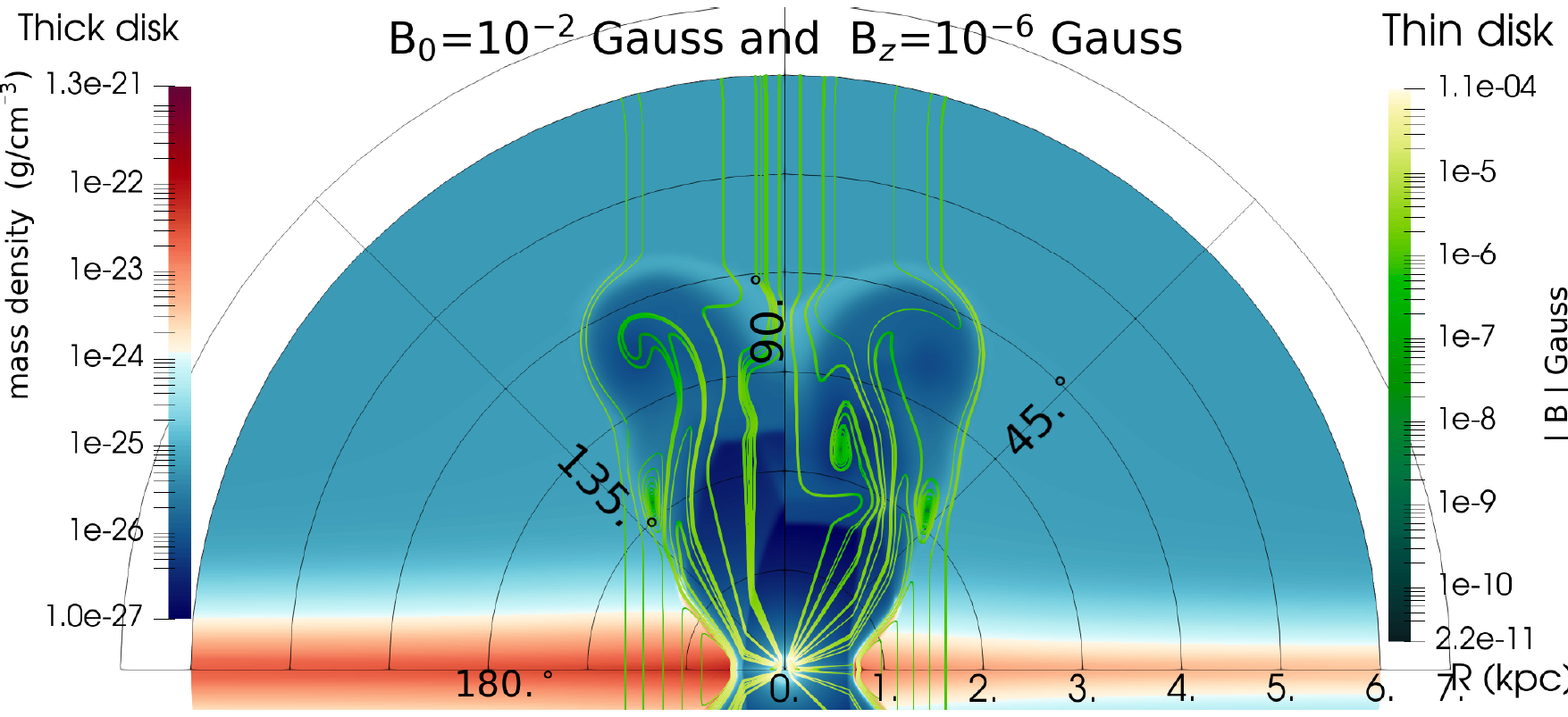}

\caption{Density contour for an expanding galactic bubble in a magnetized halo-disk medium at t=10 Myr. Each panel displays two morphologies of the disk under investigation: a thick disk on the left and a thin disk on the right. The galactic wind has a mass flux of $\dot{M}$= 1 M$_{\odot}$/yr. Top-left panel: dipole B= $10^{-4}$~Gauss, Top-right panel: dipole B= $10^{-2}$~Gauss, bottom-left panel: dipole B= $1$~Gauss, bottom-right panel: a dipole component with (B$_{0}=10^{-2}$~Gauss and uniform field B$_{\rm}=10^{-6}$~Gauss.)}
\label{fig:MHD_dipoles}
\end{figure*}


\subsection{Magnetized halo-disk model}
In the following sections, we investigate the evolution of the starburst-driven wind with a scenario that incorporates both the galactic halo-disk and the large-scale magnetic field. Specifically, we explore two configurations: thick and thin disk with a dipolar magnetic field and a uniform poloidal magnetic field.

\subsection{Cases with dipolar magnetic field $B_0 = 10^{-4}$ Gauss}

In cases where a magnetic dipole field of $B_0 = 10^{-4}$ Gauss is present, the strong galactic wind pushes the magnetic field outward, compressing it between the termination and forward shocks. In the free galactic wind region, the magnetic field exhibits a split-monopole shape.

In the region of the galactic disk, the shocks are very strong, with a significant compression ratio. Specifically, the termination shock has a compression ratio $\mathcal{R}_{\text{ts}} \approx 6$, while the forward shock has $\mathcal{R}_{\text{fs}} \approx 10$. These shocks are quasi-stationary.

At the termination shock, the magnetic pressure becomes an order of magnitude lower than the thermal pressure. However, the resulting magnetic force, which includes magnetic pressure and tension, becomes sufficient to change the properties of the shocks. Furthermore, the strong interaction between the wind and the galactic disk creates a dense shell of shocked disk material compressed by the ram pressure of the wind, extending to higher altitudes. This dense shell forms an angle of approximately $\theta \approx 40^{\circ}$ in the case of the thin disk and $\theta \approx 30^{\circ}$ in the case of the thick disk, with respect to the polar axis.

Due to the high density of the shell, $n \approx 1000 \, \mathrm{cm^{-3}}$, radiative cooling is efficient, and the temperature drops to around $T \approx 1000 \, \mathrm{K}$, resulting in thermal pressure becoming lower than the pressure of the external medium. The shell is mainly confined by the pressure of the external medium and supported from collapsing toward the galactic axis by the galactic wind. The magnetic tension force from the compressed magnetic field in this shell acts toward the polar axis and also contributes to the collimation of the shell, limiting its expansion outward from the galactic axis compared to non-magnetized cases.

This shell forms a conical wall that confines the wind and, consequently, the galactic bubble. The outflow open angle remains constant during the 10 Myr of evolution that we have followed. As the flow outruns the dense surrounding shell in height, it takes on a conical shape.

The galactic wind that interacts with the inner disk near the equatorial region is redirected along the polar axis. Figure (\ref{fig:MHDB01e_4_stream_lines}) illustrates the streamlines that trace this flow pattern. The main difference between the cases of thin and thick galactic disks lies in the region of free galactic wind. In the case of the thick disk, it is mainly cylindrical, while it is conical in the case of the thin disk.

At higher altitudes, the initial dipolar magnetic field is also compressed between the termination shock and the forward compression wave, resulting in magnetic pressure and tension that slow down the expansion of the termination shock along the galactic axis. Additionally, the wind material from the disk level that is deflected along the external dense shell, as described earlier, reaches the top of the bubble and increases the mass swept-up by the wind. This also contributes to slowing down the termination shock. As a result, the termination shock has a compression ratio $\mathcal{R}_{\text{ts}} \approx 4$. It expands slowly with a speed in the lab frame of $v_{\text{ts}} \approx 24 \, \mathrm{km/s}$, corresponding to a fast magnetosonic Mach number in the wind frame of $\mathcal{M}_{\text{ts, fast}} \approx 11$.

Regarding the forward shock, it becomes weak when the bubble starts to expand outside the galactic disk, changing from a shock wave to only a compression wave. At an age of 10 Myr, its compression ratio is $\mathcal{R}_{\rm fs}\approx 1.4$. It propagates with a speed of 125 km/s, corresponding to a fast magnetosonic Mach number of $\mathcal{M}_{\rm fast}\approx 0.1$ and an Alfvén Mach number on order of $\mathcal{M}_{\rm alfven}\approx 1$. Beyond the forward shock, the magnetic field retains its dipolar structure.

\subsubsection{Cases with dipolar magnetic field $B_0 = 10^{-2}$ Gauss}

The density contour and magnetic field for cases with a dipolar magnetic field of $B_0 = 10^{-2}$ Gauss are shown in Figure~(\ref{fig:MHD_dipoles}, top right panel). The presence of a dipole magnetic field with an amplitude of $B_0 = 10^{-2}$ Gauss decelerates the wind near the galactic disk. Consequently, the interaction between the wind and the galactic disk is weakened, resulting in thicker, less dense, and more stable shock structures and collimated shells. In contrast, for a dipole field with $B_0 = 10^{-4}$ Gauss, the resulting structures would be thinner, denser, and more susceptible to energy loss through radiative cooling.

As a consequence, the shocked shell evolves at larger distances of $0.6$ kpc (thin disk case) and $0.5$ kpc (thick disk case) from the galactic center, instead of $0.4$ kpc (thin disk case) and $0.25$ kpc (thick disk case) in the case of a $10^{-4}$ Gauss dipole field.

In the thin disk case, the termination shock retains the same conical shape as in the $10^{-4}$ Gauss cases. However, in the case of a thick galactic wind, it transitions from a conical to an elliptical shape. Near the galactic axis, in the case of a thin disk, the termination shock has a speed of $20$ km/s in the lab frame and a compression ratio of $\mathcal{R}_{\rm ts} \approx 3$. In the case of a thick galactic disk, the termination shock has a speed of $30$ km/s in the lab frame, with a fast magnetosonic number on the order of one, and a compression ratio of $\mathcal{R}_{\rm ts} \approx 3.5$.

In the case of an expanding bubble in the galactic medium with a thick galactic disk, after the galactic wind passes through the termination shock, it is accelerated by the funnel induced by the shocked shell as it expands radially. The re-accelerated wind then interacts with the edge of the shell, inducing the development of a new shock propagating toward the galactic axis. This new shock is weaker than the termination shock, with a speed of about $1$ km/s, a fast magnetosonic Mach number of $\mathcal{M}_{\rm fast} \approx 2$, and a compression ratio of $\mathcal{R}_{\rm rs} \approx 2$. Once again, the magnetic field is compressed and redirected toward the galactic axis, increasing the collimation of the flow. This new shock focuses the galactic wind and makes it less prone to vortices, contrasting with the hydrodynamic case and the case with a dipole magnetic field of $10^{-4}$ Gauss. Consequently, the galactic bubble expands at higher altitudes. It is important to note, however, that these two shocks in the galactic wind eventually merge at an age of 12 Myr.

Regarding the forward shock, in the case of a thick galactic disk, it is fast with a speed on the order of $v_{\rm fs} \approx 200$ km/s, and its compression rate is low $\mathcal{R}_{\rm fs} \approx 1.6$.

In the case of a thin galactic disk, at the age of 10 Myr, the bubble expands with a velocity of $v_{\rm fs} \approx 200$ km/s. It remains sub-magnetosonic but super-Alfvénic. At the front of the galactic bubble, there is a compression wave that is expanding, not a shock. As mentioned earlier for the unmagnetized scenario, the opening angle of the galactic bubble is larger in the case of a thin galactic disk ($40^\circ$) compared to a thick galactic disk ($35^\circ$). This difference can also explain why the bubble extends to higher latitudes when the disk is thick.

\subsubsection{Cases with a dipole of 1 Gauss}

In the case of a magnetic field dipole with an amplitude of $1$ Gauss (Figure~\ref{fig:MHD_dipoles}, bottom left panel), the magnetic field is strong and significantly influences the dynamics of the galactic bubble in both thin and thick disk scenarios.

In particular, in both cases, the magnetic field dipole directs the galactic wind primarily toward the disk. This results in a substantial deformation of the bubble and a deceleration of the expanding bubble, leading to a weak forward-compressing wave. Consequently, the termination shock advances at a speed of only $1$ km/s.

\subsection{Two-component magnetic field}

In the following scenario, we consider an external medium with a disk-halo profile featuring a dipolar magnetic field of strength $B_0 = 10^{-2}$ Gauss, along with a uniform poloidal field of strength $B_{\rm z} = 10^{-6}$ Gauss (see Figure~\ref{fig:MHD_dipoles}).

The combination of a dipolar magnetic field and a uniform poloidal magnetic field, as depicted in the previous cases, plays a crucial role in the formation of galactic bubbles. The uniform vertical component of the magnetic field enhances elongation by increasing magnetic tension and pressure, pushing the shocked galactic medium toward the galactic axis. Conversely, the dipolar magnetic field affects the flow of stellar material by slowing it down. Moreover, unlike the uniform field, the strength of the dipolar field decreases with distance as $R^{-3}$, leading to a diminishing influence as the bubble expands.

In the region of the free galactic wind, the dynamics are dominated by the wind itself, and the magnetic field is predominantly radial. However, as the bubble expands, the uniform magnetic field component begins to influence the lateral expansion, introducing non-radial effects. Despite its small amplitude, the uniform poloidal component of the magnetic field becomes dominant at large distances.

In both the thin and thick disk cases, the uniform magnetic field limits the transverse expansion of the bubble compared to the case with only a dipolar magnetic field of $B_{0} = 10^{-2}$ Gauss. In the thick disk scenario, the termination shock becomes more elongated along the polar axis, with a compression ratio of $\mathcal{R}_{\rm ts} \sim 3$. The galactic bubble also expands less radially.

\begin{figure*}[!h]
\includegraphics[width=.45\textwidth]{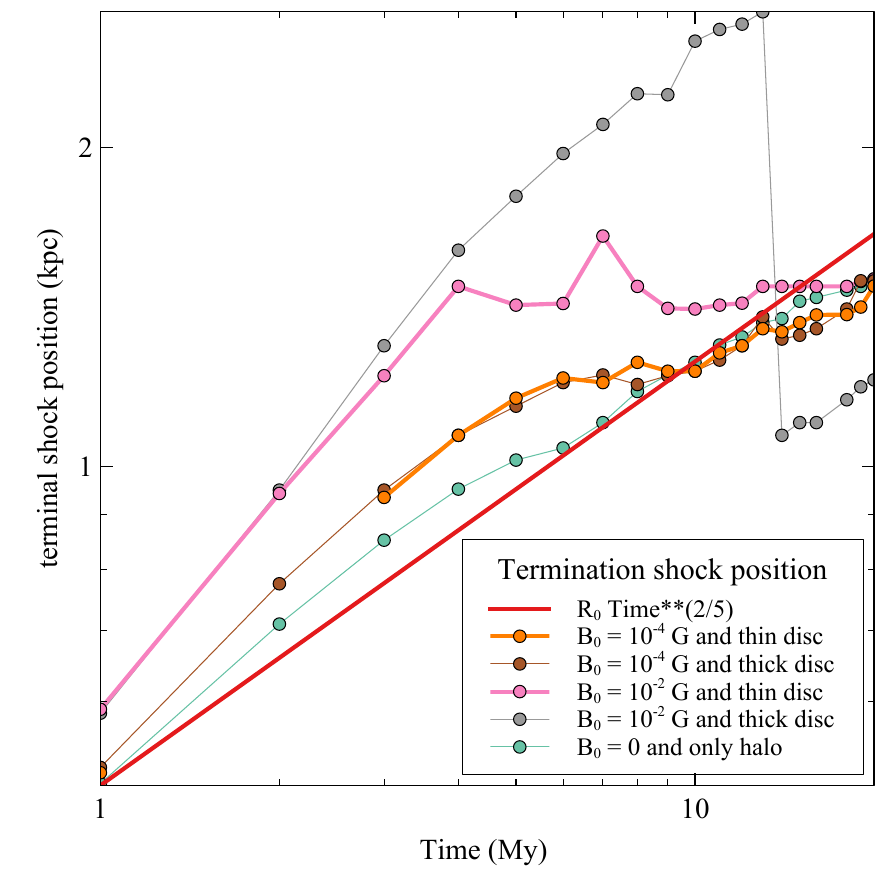}
\includegraphics[width=.45\textwidth]{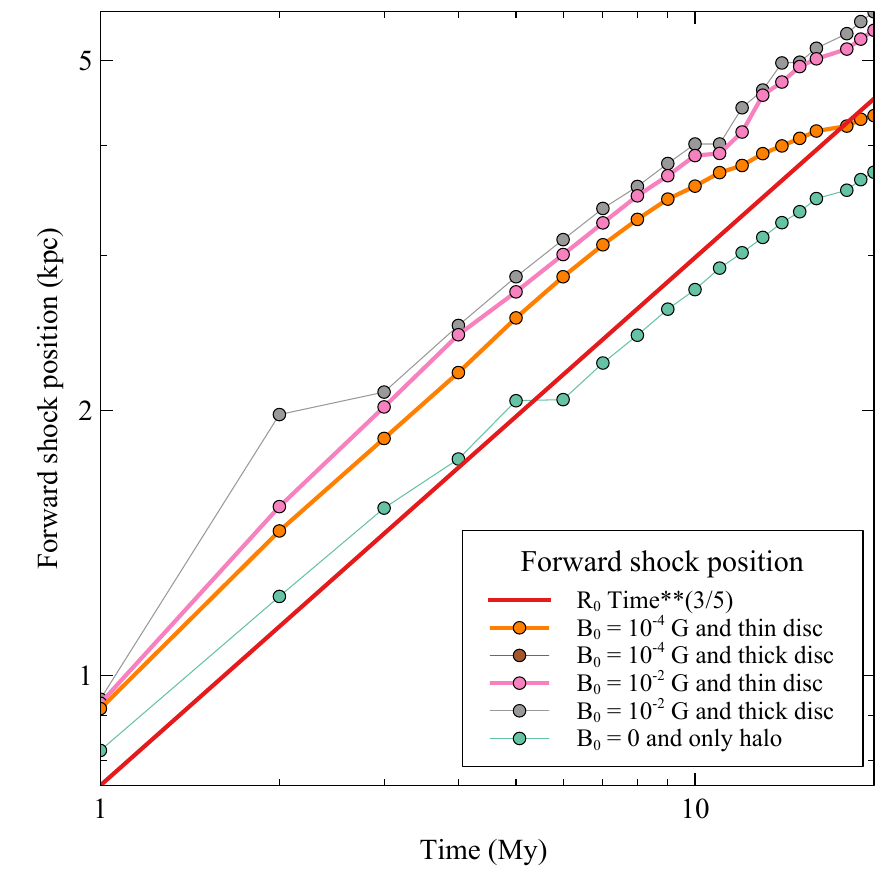}
\caption{Termination shock  and forward shock position during the evolution at an angle of 5$^\circ$ from the galactic axis for four magnetized cases and one non-magnetized case of a galactic wind bubble propagating in a halo medium. The analytical solution by \citep{weaver1977} is also shown for reference.}
\label{fig:shock_position}
\end{figure*}

\subsection{Magnetic field effects}\label{subSec:MagneticFieldEffects}

In the magnetized cases with a dipolar magnetic field and a galactic disk, we observe all three evolution phases of the termination shock. We also observe the forward shock and the compression wave near the polar axis. These observations are depicted in Figure \ref{fig:shock_position}, which illustrates the positions of the shock and the compression wave. In addition to these observations, we present an overlay of a case of an expanding galactic bubble in a halo medium (without a galactic disk). This overlay includes its corresponding fit using the analytical solution of \citet{weaver1977}.

During the initial phase of 5 Myr (for the cases with a magnetic dipole of $10^{-4}$) and 10 Myr (for the cases with a magnetic dipole of $10^{-2}$), the termination shock undergoes rapid expansion. In the second phase, the deceleration, lasting up to 12 Myr, the termination shock expands at a rate consistent with the analytical predictions of \citet{weaver1977}, specifically $R_{\rm ts}\sim t^{2/5}$. Beyond 15 Myr, the termination shock reaches a quasi-steady state, the pressure-confined phase.

The forward shock, or compression wave depending on the scenario, goes through three phases. In the first phase, it follows the expansion rate predicted by \citet{weaver1977} up to 10 Myr. After that, the forward wave slows down in the low-magnetization cases. However, in the case with a strong dipole of $10^{-2}$ Gauss, it continues to expand at the same rate. The initial phase of fast expansion of the termination shock and relatively slow forward shock corresponds to the phase of evolution of the wind inside the galactic disk, where the dipolar magnetic field has a strong influence.

In the second phase, the bubble expands into the galactic halo. In the final phase, the mass of deflected shocked wind at the galactic axis becomes sufficient to influence the dynamics of the galactic bubble.

It should also be noted that in the case of the galactic bubble expanding in the ambient medium with a galactic disk, the terminal and forward shocks along the equatorial plane exhibit quasi-steady behavior.


\subsection{Comparison with other works}\label{subSec:ComparisonWithOtherWorks}

The models studied in this paper demonstrate the effectiveness of the dense galactic disk in collimating the starburst wind at an open angle of about 45 degrees. This collimating efficiency is further enhanced when a dipolar magnetic field of magnitude $B_{0} = 10^{-2}\,\mathrm{Gauss}$ is present, resulting in a narrowing of the open angle to about 35 degrees. Additionally, considering the combined effects of a dipolar magnetic field with $B_{0} = 10^{-2}\,\mathrm{Gauss}$ and a uniform magnetic field with $B_{\rm z} = 10^{-6}\,\mathrm{Gauss}$, the open angle is reduced to about 30 degrees.

Furthermore, the models investigated in this paper show that at an age of 10 Myrs, the lateral expansion of the galactic wind at its base is between 400-800 parsecs, a dimension that scales with the solar system. This range is in agreement with the observations of the bubble base diameter as reported in \citet{Greve_2004A&A...416...67G} for the galactic bubble M82.

In all cases studied in this paper, the galactic winds exhibit a complex, multiphase, and dynamic nature. At the base of the wind, close to the inner boundary, we impose a high temperature of $5 \times 10^7$ K. This hot gas undergoes cooling due to adiabatic expansion and gradually reaches a temperature of $10^5$ K at a distance of 2 kpc from the galactic center. These observations support the findings of the galactic bubble M82, as discussed in \citet{Lopez_etal_2023ApJ...942..108L}.

However, closer to the equatorial plane, where the wind interacts strongly with the galactic disk, it is deflected at the lateral edges, leading to the formation of a hot outflow shell extending outward from the equatorial plane. In this region, the temperature drops from a high of about $7 \times 10^7$ K near the equatorial plane to $10^6$ at higher altitudes. The shell of shocked wind material is compressed both from the inside by the unshocked galactic wind and from the outside by the dense shell of the shocked cold galactic disk. As a result, it expands relatively slowly with increasing altitude, and its temperature remains greater than $5 \times 10^6$ K even when it reaches the top-hat region of the galactic bubble.

it is worth noting that radiative cooling in this region is not very efficient, mainly due to the relatively low number density, which remains at about $0.1\,\mathrm{cm}^{-3}$. Furthermore, at higher altitudes, especially near the top of the galactic bubble, the free galactic wind near the galactic axis is also heated, mainly by the termination shock.

As for the shell of the shocked galactic disk, it undergoes rapid radiative cooling, resulting in a significant temperature decrease to about $1000$ K. This cooling effect is particularly pronounced due to the high density of the shell, which is about $10^3\,\mathrm{cm}^{-3}$. It is important to note that depending on the magnetization of the galactic medium and the thickness of the galactic disk, the thickness and temperature of this shocked shell at the lateral edge of the bubble can vary. However, these variations typically remain within the same order of magnitude.

The lateral structure of the multiphase flow, with a warm galactic wind cooling with distance, the surrounding hot deflected wind, and the subsequent cold dense shocked galactic disk, differs from the results of \citet{Nguyen_Thompson_10.1093/mnras/stab2910}. In fact, we assume a spherical galactic wind emanating from the galactic center, in contrast to the shifted ring configuration studied in \citet{Nguyen_Thompson_10.1093/mnras/stab2910}. In their case, the wind converges toward the polar axis, is deflected and heated, resulting in a very high temperature at the axis.

It is noteworthy that the temperature of the wind component close to the polar axis decreases with distance until it reaches the termination shock. Meanwhile, the temperature of the hot component of the wind, which forms a shell of deflected wind, remains relatively constant with respect to the distance from the Galactic disk. This behavior is consistent with X-ray observations of the M82 galactic bubble, as suggested by \citet{Ranalli_etal_2008MNRAS.386.1464R}.

The shell of cold shocked galactic disk gas extends above the galactic disk to a height of $2\,\mathrm{kpc}$. This result is also supported by the observations of M82 by \cite{Leroy_2015ApJ...814...83L}. The behavior observed in our models is consistent with observations at multiple wavelengths and existing theoretical models \cite{Heckman_etal_1990ApJS...74..833H,
Strickland_Heckman_2007ApJ...658..258S, Leroy_2015ApJ...814...83L}.

However, since gravity is not treated in our models, there is no fountain-like flow of cold gas falling back into the galactic disk, as observed in \cite{Leroy_2015ApJ...814...83L}, which we will discuss in the next paper.

Regarding the positions of the shocks near the galactic axis (Figure~\ref{fig:shock_position}), the dipolar magnetic field accelerates the termination and forward shocks compared to the hydrodynamic case, as observed in \citet{weaver1977}, during the first 10 Myr. This acceleration increases with magnetic field strength. However, around 10 Myr, the forward shock starts to slow down due to the increased mass at the top of the bubble, consisting of the deflected wind in the region of the galactic disk.

it is essential to note that in this initial phase of the project, our focus is on investigating the influence of the galactic environment and magnetic fields on flow dynamics, bubble morphology, and shock intensities. Our model shows that the magnetic field contributes to the confinement of the galactic outflow, resulting in an increase in the expansion of the galactic bubble along the galactic axis. Consequently, in the presence of a dipole magnetic field with a strength of $B_{0} = 10^{-2} \, \mathrm{Gauss}$, the galactic bubble expands to an altitude of about 1 kpc more than in the non-magnetized case.

We have not yet considered wind variability or three-dimensional effects, which may give rise to additional instabilities and filamentary structures, as shown in previous studies such as \cite{Melioli_2013}. Our model assumes a continuous galactic disk and halo. Consequently, we have not considered the behavior of clouds embedded in this environment, which may interact with the galactic medium, undergo hydrodynamical disruption, and contribute to X-ray emission. This phenomenon has been studied extensively, both semi-analytically by \cite{Fielding_Bryan_2022ApJ...924...82F} and through 3D hydrodynamical simulations by \cite{Cooper_etal_2009ApJ...703..330C}. The dynamics and resulting emission from the interaction between the galactic wind and cold dense clumps have also been studied by \cite{Wu_etal_2020MNRAS.491.5621W}. They emphasize the importance of the charge exchange process in contributing to the total X-ray emission, supported by observations of M82 \citep{Zhang_etal_2014ApJ...794...61Z}. This process occurs at the boundary between the neutral and ionized phases, corresponding to the cool neutral gas from the clouds interacting with the hot ionized gas in the galactic outflow.


\section{Multiwavelength and multi-messenger implications}

\subsection{Thermal X-ray emission}

To distinguish between the models, we conduct a morphological comparison of the soft X-ray emission originating from the hot gas filling the bubble, specifically the shocked wind and the layer of shocked surrounding medium. We generate X-ray emission maps using the results of our numerical simulations. We focus on gas with a temperature >$10^3$ K, assuming it is fully ionized, and calculate the X-ray emission coefficient by solving,

\begin{equation}
    j_{v}=n_{\rm{e}}^2 \cdot \chi(T,Z) \,,
\end{equation}

where $n_{\rm{e}}$ is the electron density, $T$ is the gas temperature, $Z$ is the metallicity, and $\chi(T,Z)$ is the emissivity coefficient. The latter is obtained by fitting the \textsc{CHANTI} database for solar metallicity in the soft X-ray range from 0.2 to 2 keV (see \citealt{dere1997}).


\begin{figure}[h!]
    \centering
    \includegraphics[width=0.35\textwidth,angle=-90]{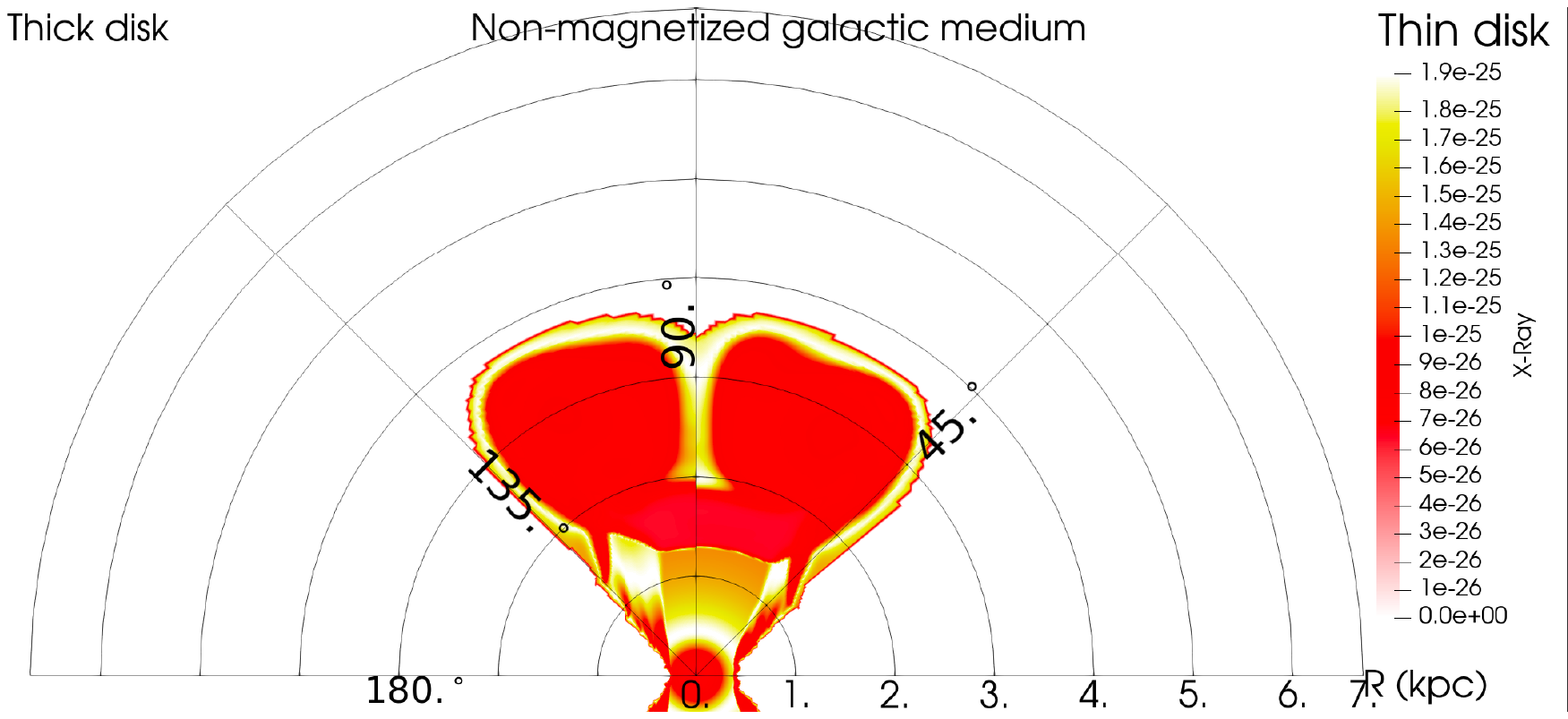}
    \includegraphics[width=0.35\textwidth,angle=-90]{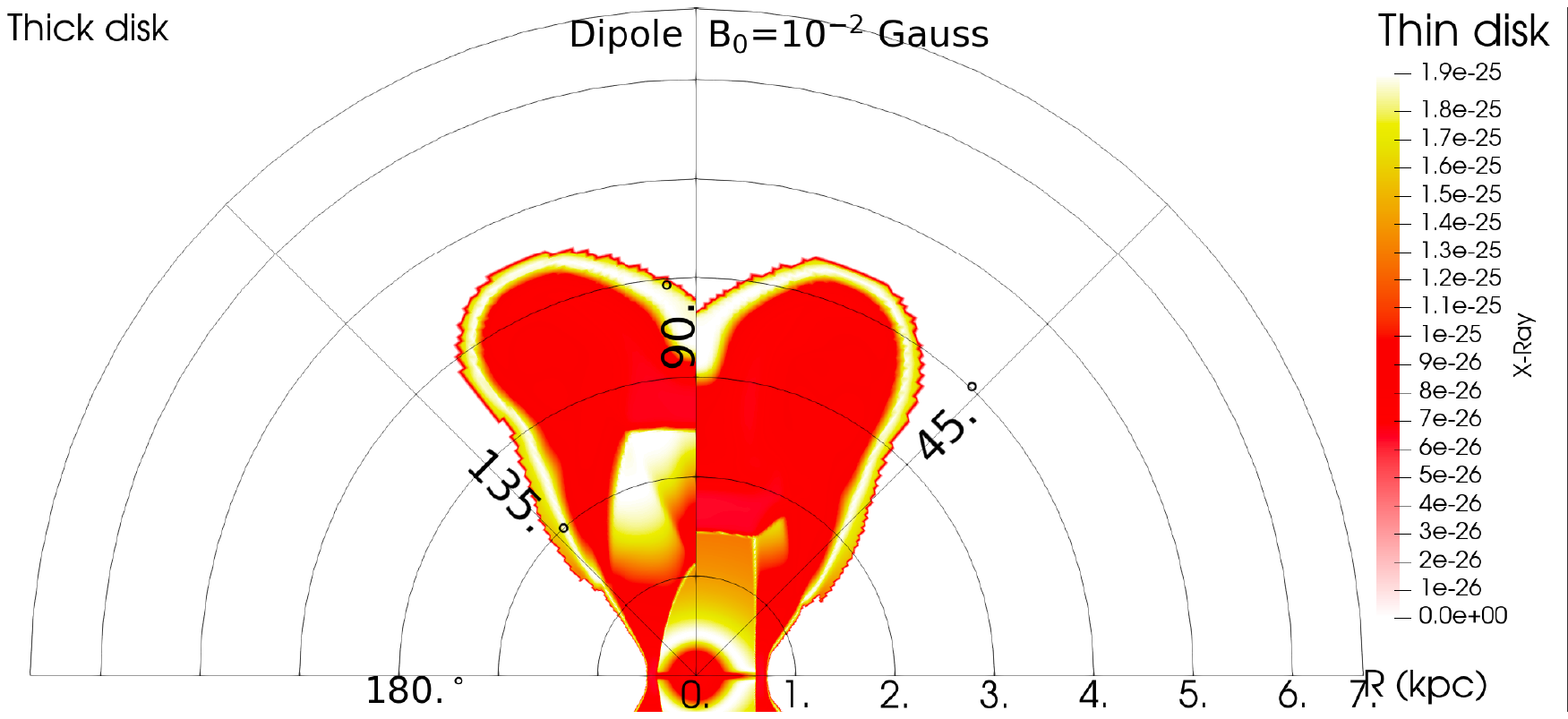}
    \includegraphics[width=0.35\textwidth,angle=-90]{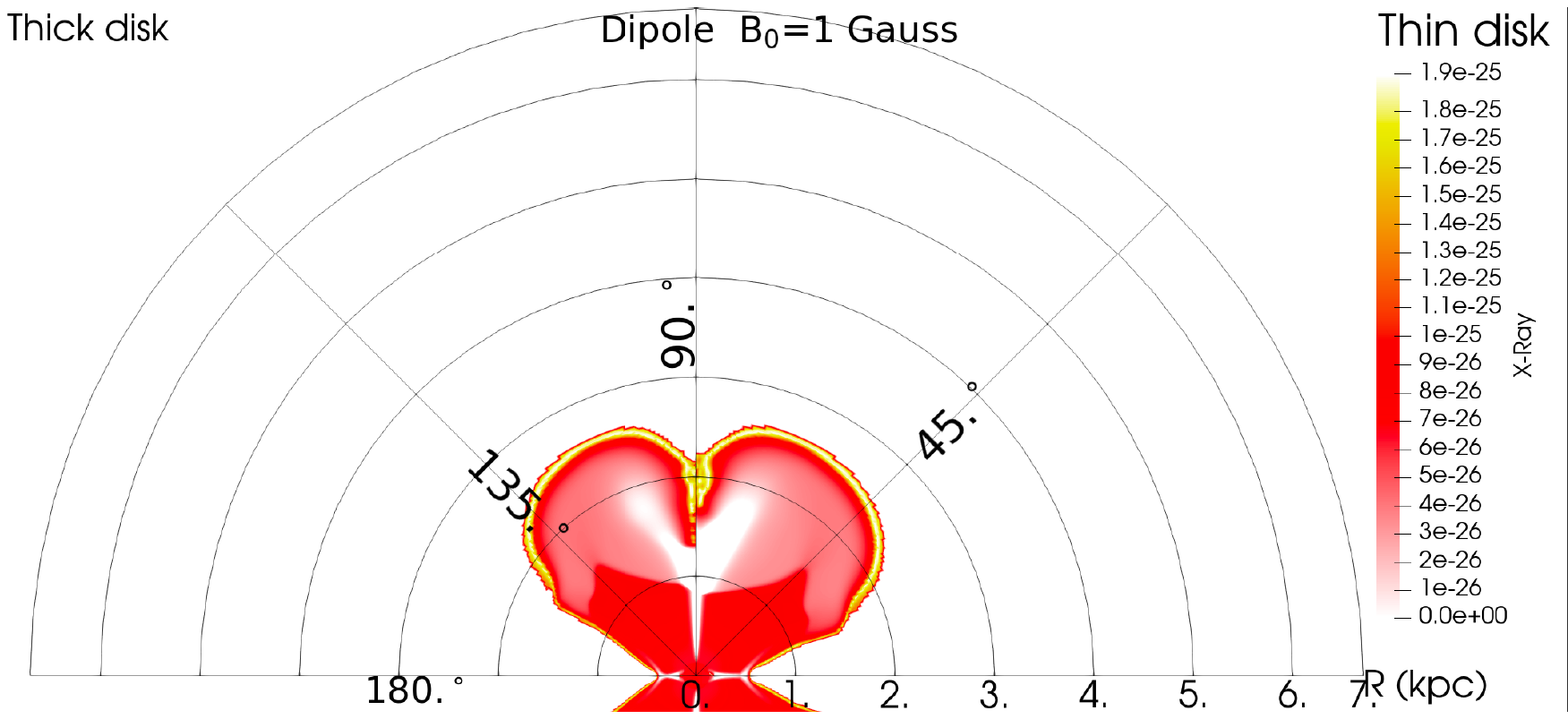}
    \caption{Soft X-ray emission maps for models A-B (non-magnetized cases), upper panel, models K-L (magnetized galactic medium with a dipole $B_{0}=10^{-2}$ Gauss), medium panel, and model M-N, lower  panel (magnetized galactic medium with a dipole $B_{0}=1$ Gauss) }
    \label{fig:xray_mab}
\end{figure}

In models A and B (unmagnetized models, top panel of Fig. \ref{fig:xray_mab}), a significant portion of the X-ray emission originates from the gas heated by the termination shock and the forward shock/compression wave. In these regions, the temperature ranges from $10^6$ to $10^7$ K. At the lateral edge of the bubble, the termination shock primarily dominates the X-ray emission. At higher altitudes, however, the emission from the forward shock/compression wave becomes dominant. Additionally, in this zone, behind the forward front, the presence of turbulent shocked wind contributes to the X-ray emission. However, the free galactic wind contributes to the X-ray emission with a decrease with altitude as its temperature decreases due to adiabatic expansion until it reaches the termination shock, where the X-ray emission increases again.

Furthermore, surrounding the free wind region is a cone of gas that has cooled due to rapid expansion. This region shows weaker X-ray emission.

Disk Dipole $10^{-2}$ G
In the case of a disk (thick and thin) with a magnetic field of $10^{-2}$ Gauss (models K and L, middle panel of Figure \ref{fig:xray_mab}), a structure similar to the purely hydrodynamic models can be observed, although the post-shock gas emission at the front may not be as intense as in the HD models. In these cases, the emission from the region of shocked injected gas is significantly enhanced. This enhancement is primarily due to the presence of a strong magnetic field, which introduces a higher level of complexity to the deflected wind at the equatorial plane induced by the galactic disk. This interaction between the deflected wind and the nearby magnetic field leads to the formation of two shocks within the galactic wind. As a result, the vertical extent of the shocked region increases, further contributing to the higher emission observed in these cases.

Similar to the models without a magnetic field, the X-ray emission can be observed at heights exceeding 50 or 60 times the height of the disk.
However, the models with the highest magnetic field intensity (models M and N) show a much more compact region of soft X-ray emission of about 10 times the height of the disk.

It is important to note that observations of galaxies with intense star formation, such as M82, show that the X-ray emission extends to significant heights above the disk, exceeding the height scale by more than 60 times. This emission appears to come from the superbubble. The emission in question could be caused by gas injected by the formation flare. For example, supernova explosions or instabilities could encounter the reversing shock wave, which interacts with the gas injected by the superbubble nebula.

\subsection{Particle acceleration and transport}

The magnetic and geometrical properties of starburst-driven wind bubbles are crucial for the acceleration and transport of high-energy particles. In our simulations, the presence of strong shocks in the system implies efficient diffusive shock acceleration.

Regardless of the disk-halo configurations, the outflow does not exhibit strong shock conditions at the forward front. The typical Mach number found at the forward front is of order unity, $\mathcal{M}_{\rm fs} \approx 1$. On the other hand, the wind termination shock usually leads to the formation of strong shocks with large compression ratios in the range of approximately 3.6-6, associated with a Mach number greater than 1.

Additionally, we observe that radiative cooling in the shocked wind is modest, while the dominant cooling occurs in the shocked ambient medium near the equatorial plane. These results align with the well-known analytical theory of wind-blown bubbles \citep{weaver1977, koo1992}, where we clearly observe the free expansion, deceleration, and pressure-confined phases.

In agreement with what has been proposed in \citep{Peretti2022}, we find that the physical conditions met at the wind termination shock are ideal for particle acceleration by diffusive shock acceleration. Specifically, assuming Bohm diffusion, it is possible to estimate an absolute upper limit for the maximum energy by equating the upstream diffusion length to the termination shock radius. This gives a typical maximum energy:

\begin{equation}\label{Eq:Electric_field_cr}
    \frac{E_{\rm max}}{Z} \lesssim 1.9 \cdot 10^2 \, \rm PeV \, \left( \frac{B}{10 \, \mu G} \right) \left( \frac{v_{\rm wind}}{2000 \, km \, s^{-1}} \right) \left( \frac{R_{\rm ts}}{  kpc } \right), 
\end{equation}

where $Z$ is the electric charge of the cosmic ray species, $B$ is the local magnetic field at the shock, and $v_{\rm wind}$ is the terminal wind speed, which we use as a proxy for the upstream plasma velocity in the shock reference frame, since $v_{\rm ts} \ll v_{\rm wind}$.
This suggests that starburst-driven winds could contribute substantially to populating the cosmic ray spectrum between the knee, $\sim$ 3 PeV, and the ankle, $\sim$ 3 EeV. It would thus emerge as one of the most promising candidate cosmic ray sources responsible for the transition from galactic to extra-galactic origin.

It is important to note that Bohm diffusion provides a model-independent upper limit on the true maximum energy. A more realistic estimate would require the assumption of a specific type of turbulence with the associated coherence length. This consideration is beyond the scope of our work. Nevertheless, we note that the Larmor radius of the highest energy particles can be found in the range of $10-10^2$ pc, assuming $B \approx 10 \, \text{G}$. Such a size is compatible with the typical expansion of starburst nuclear regions and thus with a reasonable estimate on the order of magnitude of the coherence length. Since the Larmor radius is expected to be in the same order of magnitude as the coherence length, we note that the actual maximum energy of the particles may be relatively close to our upper limit. For further discussion, see \citep{Peretti2022}.

In contrast to simple analytical models, we observe that the presence of the disk and large-scale magnetic fields can strongly affect the spherical expansion and lead to a collimation of the outflow. This can have a significant impact on the termination shock geometry, leading to a situation close to "plane parallel." 
In this context, the adiabatic energy losses in the system should be limited while taking into account the lateral escape. Such a geometric difference from a pure spherical geometry should not have a major impact on the maximum energy. However, if relevant, the lateral escape from the wind bubble could lead to diffusion in the galactic halo. A significant fraction of the accelerated particles would then be able to diffuse back to the galactic disk, leading to an efficient production of PeV neutrinos and gamma rays.

\section{Discussion and conclusions}
In our study, we performed MHD simulations with the AMRVAC code to investigate the evolution of a galactic wind from a typical starburst nucleus. 
Our focus was particularly on understanding the effects of the  large-scale magnetic field and dense galactic disk on the evolution of the large-scale wind-blown bubble, which extends up to kiloparsec  scales. 
The structure and strength of the magnetic field in the halo where the starburst  wind expands are not yet well constrained and, in this paper, we considered various strengths and two different geometries -- a regular parallel field and a dipole -- to support our discussion.

Our work has revealed that the presence of a magnetic field can have a significant impact on the formation and characteristics of SBNs. When the magnetic field is associated with the galactic disk, it induces modifications in the shape of the SBN and can even affect the strength of the shocks within it. Specifically, the forward shock may undergo a transition into a compression wave, resulting in the loss of its strong shock jump conditions. On the other hand, the wind termination shock often maintains its strong shock properties, making it an ideal location for the process of diffusive shock acceleration. This process has the potential to generate particle energies reaching up to $10^2$ PeV.
Exploring the parameter space (stellar's mass flux, galactic medium structure and its magnetization), we found that the structure of the magnetic field alone, unless rooting for very specific tuning, does not naturally lead to the shaping of the large-scale bubbles as observed in the case of M82. 

Our findings, considering the presence of a galactic disk, indicate that a dipolar structure of the large-scale magnetic field has a significant influence on the shaping of wind bubbles and shock surfaces. Interestingly, the magnetic field weakens the forward shock of the bubble, transforming it into a compression wave. Furthermore, we observe that the magnetic field exhibits a monopolar structure within the region of free galactic wind, while displaying a more turbulent structure at higher altitudes within the galactic bubble. This turbulence is a result of the reflected shocked wind at the galactic level propagating upward and inducing vortices at high altitudes within the shocked bubble. 

The turbulent structure of the flow and magnetic field in the top hat of the galactic bubble exhibits remarkable spatial extension, extending up to 2 kpc in the vertical direction and 1 kpc in the radial direction, even at an age of 10 Myr. An interesting feature within this region is the presence of a large-scale toroidal  electrical current, which undergoes directional changes.
In the case of a dipolar magnetic field in the shocked region, the vortex generated by the turbulent flow induces an inversion in the toroidal electric field. Specifically, the magnitude of the toroidal electric field is approximately $E \approx -\frac{(v\times B)}{c}\approx 10^{-7}$ N/C (CGS units). This distinctive behavior of the magnetic field, characterized by the inversion of the toroidal electric field,  could play an interesting role for particle acceleration.

In addition, we have generated X-ray emission maps as a means to compare our numerical results with observed properties of starburst galaxies. 
Remarkably, our models incorporating a dipolar magnetic field with magnitudes less than 10$^{-2}$ Gauss exhibit soft X-ray emission in regions that are shocked by the forward and termination shocks, at distances of 50 to 60 scales of the height of the galactic disk.  
It is important to highlight that despite the initial dipole magnetic field strength being $B_0 = 10^{-2}$ G, the magnetic field within the region of the free galactic wind maintains a relatively weak strength, on the order of $4\;\mu$Gauss at a distance of 1 kpc. In the free wind region, the magnetic field does indeed get stretched.
This X-ray emission pattern closely resembles the observed emission in similar objects. Notably, the X-ray emission also reveals an extended envelope surrounding the galactic wind, spanning up to 2 kpc, which is consistent with the structures observed in the study by \cite{lopezrodriguez2021}.

As a continuation of our research, we are currently exploring the influence of stellar evolution and cosmic rays on the evolution of galactic bubbles.   We aim to gain a more comprehensive understanding of the complex interplay between different physical processes in the evolution of galactic bubbles, and how these processes impact the overall dynamics and properties of galactic winds. 
\begin{acknowledgements}
We thank the anonymous reviewer for constructive feedback that improved the manuscript. 
PC acknowledges funding from the European Union's Horizon 2020
research and innovation programme under the Marie Sklodowska-Curie grant agreement No 945298 ParisRegionFP. 
Computations of the MHD results were carried out on the within EDARI allocation \texttt{A0090406842} and \texttt{A0100412483}) and  the MesoPSL cluster at PSL University\footnote{\url{http://www.mesopsl.fr/}} in the Observatory of Paris. This work was granted access to the HPC resources of MesoPSL financed by the Region Île-de-France and the project Equip@Meso (reference \texttt{ANR-10-EQPX-29-01}) of the programme Investissements d’Avenir supervised by the Agence Nationale pour la Recherche.
The research activity of EP was supported by Villum Fonden (project No.~18994) and by the European Union’s Horizon 2020 research and innovation program under the Marie Sklodowska-Curie grant agreement No. 847523 "INTERACTIONS". EP was also supported by Agence Nationale de la Recherche (grant ANR-21-CE31-0028).
\end{acknowledgements}


%
%

\bibliographystyle{aa.bst} 
\bibliography{biblio.bib}

\end{document}